\documentclass[preprint2]{aastex}

\shorttitle{Extrasolar planet taxonomy}
\shortauthors{Marchi S.}

\begin{document}


\title{Extrasolar planet taxonomy: a new statistical approach}

\author{Simone Marchi}

\affil{Dipartimento di Astronomia, Universit\`a di Padova\\ vicolo
  dell'Osservatorio 2,  35122 Padova, Italy}
\email{simone.marchi@unipd.it}









\begin{abstract}

In  this paper  we present  the  guidelines for  an extrasolar  planet
taxonomy.  The discovery of an increasing number of extrasolar planets
showing a  vast variety of {\it planetary  parameters}, like Keplerian
orbital  elements  and {\it  environmental  parameters}, like  stellar
masses, spectral types, metallicity etc., prompts the development of a
planetary  taxonomy.  In  this work  via principal  component analysis
followed by hierarchical clustering analysis, we report the definition
of  five robust  groups  of  planets.  We  also  discuss the  physical
relevance  of such  analysis,  which  may provide  a  valid basis  for
disentangling the role of  the several physical parameters involved in
the  processes  of planet  formation  and  subsequent evolution.   For
instance, we were able to divide the hot Jupiters into two main groups
on the basis of their stellar masses and metallicities.  Moreover, for
some  groups,   we  find  strong   correlations  between  metallicity,
semi-major axis and eccentricity.   The implications of these findings
are discussed.

\end{abstract}

\keywords{ planetary systems --- planetary systems: formation ---
 planets and satellites: general   }

\section{Introduction}

With  the discovery  of the  first extrasolar  planet (EP)  orbiting a
solar-like star  \citep{may95} a  new field of  astronomical research
started.   So  far,  more   than  200  extrasolar  planets  have  been
discovered.   Despite  severe  observational  biases  -only  partially
overcome  by  the development  of  more  sophisticated techniques  and
facilities- the most striking fact is the vast variety of EPs, and the
remarkable difference with respect to the planets of our Solar System.
Traditional  theories  of  planetary  formation  have  been  seriously
challenged by these discoveries, and there are many aspects that still
need   to  be   understood  in   detail,  like   planetary  migrations
(disk-embedded vs planet-planet interactions) and the influence of the
metallicy on the planetary  formation processes, just to mention some.
The heterogeneity of EPs has been analyzed in several papers, and some
important trends have  been uncovered \citep[e.g. see][]{zuc02, san03, udr03,
egg04, soz04, fis05}.  
On the other  hand, planetary formation
is a very complex process, where a number of parameters have important
effects, as shown by several theoretical works.
It is also likely that some parameters act simultaneously in a complex
way, thus  motivating the need  for a multidimensional  approach going
beyond the search for  simple correlations among individual parameters.
In  other   words,  the  planetary  formation  process   occurs  in  a
multidimensional space. In addition  to Keplerian orbital elements and
stellar properties, other relevant parameters may be discovered in the
future, as the formation theories improve.\\
We adopt a multivariate approach to EPs in order to uncover underlying
trends  which  may   provide  important  information  about  the planet
formation processes.   The first step  is the development of  a robust
taxonomy for EPs.  Taxonomy, just  as in other fields of research, may
be a  precious tool  in defining  clusters of EPs,  which in  turn may
highlight  differences  in  the  formation  processes  and  subsequent
evolution.\\
%
This will be of particular importance with the increasing number of EP
discoveries which  is expected to occur  in the near  future thanks to
several  space missions  (e.g.   CoRoT and  Kepler) and  ground--based
surveys.  For  the moment, the number  of EPs is about  200, which may
prevent firm  statistical conclusions.  Nevertheless,  here we propose
the guidelines for  an EP taxonomy, to be refined  as more data become
available.\\
In  the next section  (2), we  describe the  parameters used  for this
study  and  we  perform  a   statistical  approach  of  the  data  for
dimensional scaling  purpose.  Then, in section  (3), cluster analysis
is  performed.  We  caution from  the beginning  that  these analyses
-much  like all statistical  analyses- are  not unique,  since several
different criteria may be used  for characterizing the data.  There is
no a  priory best criterion: the  choice can indeed  vary according to
the  kind of  data under  analysis.  We  shall provide  a step-by-step
justification for all the choices  made.  Finally, in section (4), the
solution will be discussed along with its physical interpretation.

\section{Multivariate analysis and dimensional scaling of EPs}

\begin{figure}
\includegraphics[scale=.40]{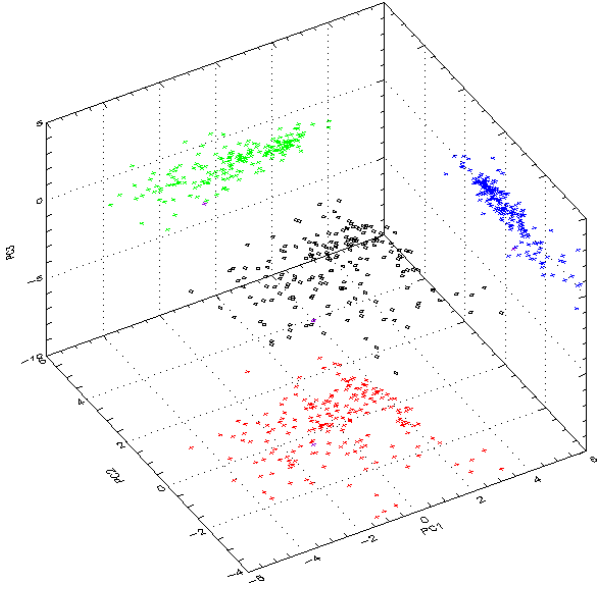}
\caption{Distribution of the data in the 3-fold clustering space.}
\label{pca3d}
\end{figure}

The inputs to  our model are the elements  provided by the interactive
extrasolar       planets      catalog       mantained       by      J.
Schneider\footnote{http://exoplanet.eu/}.     These   are:   planetary
projected mass  ($M_p$), orbital  period ($P$), semi-major  axis ($a$),
eccentricity  ($e$),  inclination  ($i$),  stellar mass  ($M_s$),  and
stellar  metallicity ([Fe/H]).   Other possibly  important parameters,
like stellar spectral  type and stellar age, have  not been considered
at  this stage,  but will  be  the subject of  further refinements.   Only
objects having simultaneously estimates for \{$M_p,P,a,e,M_s$,[Fe/H]\}
have  been  used.  Notice  that  the  period  and semi-major  axis  are
obviously  correlated,  thus  only  one  of  them  is  used  (see  the
following).  We refer to them
as the  {\it input variables}\footnote{We  chose to use  simply $M_p$
and $a$ instead of performing some transformation of the data prior to
the statistical analysis, as done  by some authors which made recourse
to logarithms,  because there is  no strong physical reason  for that.
Anyway,  we  plan to  exploit  this  possibility  in further  works.}.
Therefore, each planet  can be regarded as a point  in a 5-fold space.
The following analysis  has been carried out in  the IDL language, and
it is  structured so  as to  be easily updatable  as more  data become
available.\\
We consider 183 EPs (updated at  8 November 2006).  To them, the Solar
System  planet Jupiter  has been  added.  This has  been done  because
Jupiter-like  bodies  are   approaching  the  observability  limit  in
extrasolar  planetary  systems thanks  to  recent  improvement in  the
surveys, and  also to have a  direct comparison with  our own Solar
System. We decided   not to add the other Solar  System planets as they
are still well below the detection limits.
The first step  is to perform a statistical analysis  in order to find
out if there are useless  -or less significant- input variables.  This
is  done using  principal  component  analysis (PCA).   As  this is  a
standard technique in multivariate  data analysis, details will not be
discussed here \citep[interested  readers may consult][]{eve01a}.  The
basic idea of PCA  is to combine the input variables in  such a way as
to show those of most importance.  This is done by describing the data
with a number of new variables  $pc_1 ...  pc_l$ for $l=5$, ordered in
terms of decreasing variance.  The $pc_i$  are chosen in such a way as
to  be uncorrelated  with each  other.  This  is done  in  practice by
building an $n\times l$ input data  matrix (where $n$ is the number of
planets).  From this matrix, the $l\times l$ correlation or covariance
matrix is  computed, where the correlation matrix  is more appropriate
whenever the  input variables are measured with  disparate units.  The
$pc_i$ are  eigenvectors for  those matrices.  Considering  that input
variables have different units and that some span orders of magnitudes
(e.g.   $M_p$) while  others span  a limited  range (e.g.   $M_s$), we
opted to perform variable standardization of the input variables, i.e.
to  scale input  variables  in such  a way  as  to obtain  a mean  and
variance of  1.  This  has the  advantage of allowing  the use  of the
covariance matrix,  for which the  eigenvalues ($\lambda_i$) represent
the  portion  of   the  variance  of  the  original   data  which  the
corresponding eigenvectors ($pc_i$) account for.  PCA can thus provide
a useful means for finding variables of little significance.
On the  basis of the  variance attained by  each $pc_i$ we  may reject
some  of them.  This  procedure has  the advantage  of using  only the
variables which  are important, allowing a simpler  description of the
data set with only a minor loss of information.\\
As  a first step,  we used  PCA to  decide whether  the period  or the
semi-major axis is more suitable for the statistical analysis (see also
\S   3.2   on  this   point).   PCA   was   then  performed   on   the
\{$M_p,P,e,M_s,$[Fe/H]\}  and  \{$M_p,a,e,M_s,$[Fe/H]\}  spaces.   The
choice was made by requiring that the variance of the resulting $pc_i$
is more  concentrated in the  first principal components.   The 5-fold
\{$M_p,a,e,M,{\rm  [Fe/H]}$\} space  turned  out to  be slightly  more
appropriate.  Now, no commonly accepted rules exist for deciding which
variables -if any- can be  safely thrown away.  One criterion suggests
keeping  variables which  account for  70-90\% of  the  total variance
\citep{eve01a}.
Thus  it seems  reasonable  to  keep only  the  first three  principal
components which  account for  73\% of the  total variance.   The fact
that  we  can use  only  3  principal  components allows  a  graphical
representation  and  a  better  control  of  the  results.   Moreover,
dimensional  scaling  has the  effect  of  simplifying the  clustering
analysis. This  is why  we employed PCA,  but we  are aware it  is not
strictly   necessary  for   the   development  of   a  taxonomy.    In
fig. \ref{pca3d} the  first three principal components are shown.\\
To the accepted levels of  variance we  may write the  following {\it
decomposition} formulae:

\begin{eqnarray*}
pc_1 & = & -0.206\cdot (M_p-2.64)-0.543\cdot (a-1.19)-\nonumber\\
 & & 3.417\cdot (e-0.25)- 1.823\cdot (M_s-1.03)+\\
 & & 0.067\cdot({\rm [Fe/H]-0.1})\\
pc_2 & = & -0.083\cdot (M_p-2.64)-0.030\cdot (a-1.19)-\nonumber\\
 & & 0.335\cdot (e-0.25)+3.080\cdot (M_s-1.03)+\\
 & & 3.921\cdot({\rm [Fe/H]-0.1})\\
pc_3 & = & 0.090\cdot (M_p-2.64)-0.329\cdot (a-1.19)+\nonumber\\
  & & 1.795\cdot (e-0.25)-1.748\cdot (M_s-1.03)+\\
  & & 2.063\cdot({\rm [Fe/H]-0.1})
\end{eqnarray*}





where each $pc_i$ is expressed in terms of a linear combination of all
the  input variables.   Notice  that [Fe/H]  has  little influence  on
$pc_1$,  while $a$  and  $e$  basically do  not  contribute to  $pc_2$
(consider that input variables span different ranges, so for instance,
the term containing  $e$ in equ. 2 has lower  importance than the term
involving $M_p$). These  formulae may be used to  add new planets into
the database, without  repeating the PCA (this is  useful when we want
to add only single objects  which do not alter the overall statistics,
otherwise PCA has to be repeated).\\
Finally  notice  that  we  also  checked for  the  presence  of  overall
correlations within the database. We found significant correlation (see
also  below) between  $M_p-e$,  $a-e$ and  $M_p-a$.  However,  overall
correlations  have  been widely  analyzed  elsewere  and  will be  not
discussed here \citep[e.g.][and references therein]{san05}.

\section{Cluster analsysis }

Cluster analysis is a standard technique used in a variety of research
fields from  social sciences to  geology, engineering and so  on.  The
main purpose of  such analysis is to find clusters  in a given dataset
such that elements belonging to the same cluster have a certain degree
of homogeneity,  while elements  of different clusters  have to  be as
different  as  possible.  As  with  PCA, there  is  no  unique way  of
performing  the analysis and  a number  of clustering  algorithms are
available.  The choice of  the clustering technique is quite arbitrary
and it  relies mostly on  the kind of  description of the data  we are
interested in.   When the number  of clusters is not  known a priory,
as in  our case, hierarchical  clustering is more suitable,  thus we
adopt  this technique  \citep{eve01b}.   The major  advantage of  this
technique is  that it  provides a classification  which consists  of a
series of nested partitions, which go from a single cluster containing
all objects  to $n$  clusters each containing  a single  object.  This
process is based on a  {\it proximity} parameter, usually a measure of
distance  among   objects.   Moreover,  this   classification  may  be
represented by  a two-dimensional  diagram known as a {\it dendrogram},
which  illustrates the  nested nature  of the  hierarchical partition.
However, there are a number  of possible ways to perform the analysis,
and  an accurate  step-by-step evaluation  of  the process  has to  be
performed.  In  the following we refer  to the variables  used for the
clustering (namely the \{$pc_i$\}) as {\it clustering variables}.  The
general  guidelines  for  hierarchical  cluster  analysis  are  the
following \citep[for more details see][]{eve01b}:

\begin{itemize}

\item
standardization  of  the   clustering  variables;  
\item
computation of  the  proximity  matrix,  that  is,  evaluate  the  degree  of
proximity among dataset members. This is usually done by computing the
distance   in  the   clustering  space   (in  our   case   the  3-fold
$\{pc_1,pc_2,pc_3\}$ space).  Notice that several metrics  may be used
for this purpose;
\item
definition of inter-group proximity measures. These specify the method used to
quantify  the proximity level  of two  clusters.  
Several methods are available, like  {\it single linkage}
and {\it  centroid linkage};  
\item
computation of  the distortion of the dendrogram. The dendrogram itself is a
representation  of the  original  data, where  the  separation of  two
members is specified by the  minimum distance between the two clusters
which contain  the objects.  Thus the quality  of a dendrogram  can be
evaluated by comparing the  original distance (i.e.  in the clustering
space)  of  members  with  the   distance  assigned  to  them  by  the
dendrogram.   This is  formally done  by the  so-called  {\it cophenetic
coefficient}\footnote{To  compute $\hat{c}$ proceed  as follows.   First, compute  the distance
between two  planets (indicated  by indices $i$  and $j$) assigned  by the
dendrogram. Let $d_{ij}$ be this distance, and $d$ the average value of
all $d_{ij}$.  Let $r_{ij}$ be the real distance between planet  $i$
and $j$, and $r$ the average value of all $r_{ij}$. Thus:
$$\hat{c} =\frac{ \sum_{i}\sum_{j>i}(r_{ij}-r)(d_{ij}-d) }{
  \big[\sum_{i}\sum_{j>i}(r_{ij}-r)^2\sum_{i}\sum_{j>i}(d_{ij}-d)^2\big]^{1/2} }$$} 
($\hat{c}$).  It normally ranges from 0.6 to 0.95, where
higher values correspond to a lower distortion;
\item
definition of the best partition.  In standard analysis this means stopping
the hierarchy at a given distance;  in other words to cut a dendrogram
at a  particular distance  (or {\it height},  $h$).  Again,  no unique
rule  exists to  define the  best cut.  Here we  follow  the procedure
suggested  by \cite{moj77}, the  so-called {\it  upper tail  rule}. In
detail, we first evaluate the distribution of the heights $N(h)$, then
the best cut height is determined by

$$ h_\gamma = \overline{h} + \gamma\cdot\sigma_h$$

where  $\gamma$  is  a  coefficient  which ranges  from  1.25  to  3.5
\citep{mil85}, $\overline{h}$ is the  mean of the height distribution
and $\sigma_h$ its  standard deviation.  A possible way  to define the
best $\gamma$ is to find the  values of $\gamma$ for which the maximum
separation between two consecutive solutions is achieved;
\item 
testing of the quality of the solution against the absence of clusters.

\end{itemize}

\subsection{Finding the best solution}

\begin{figure}
\includegraphics[angle=0,scale=.50]{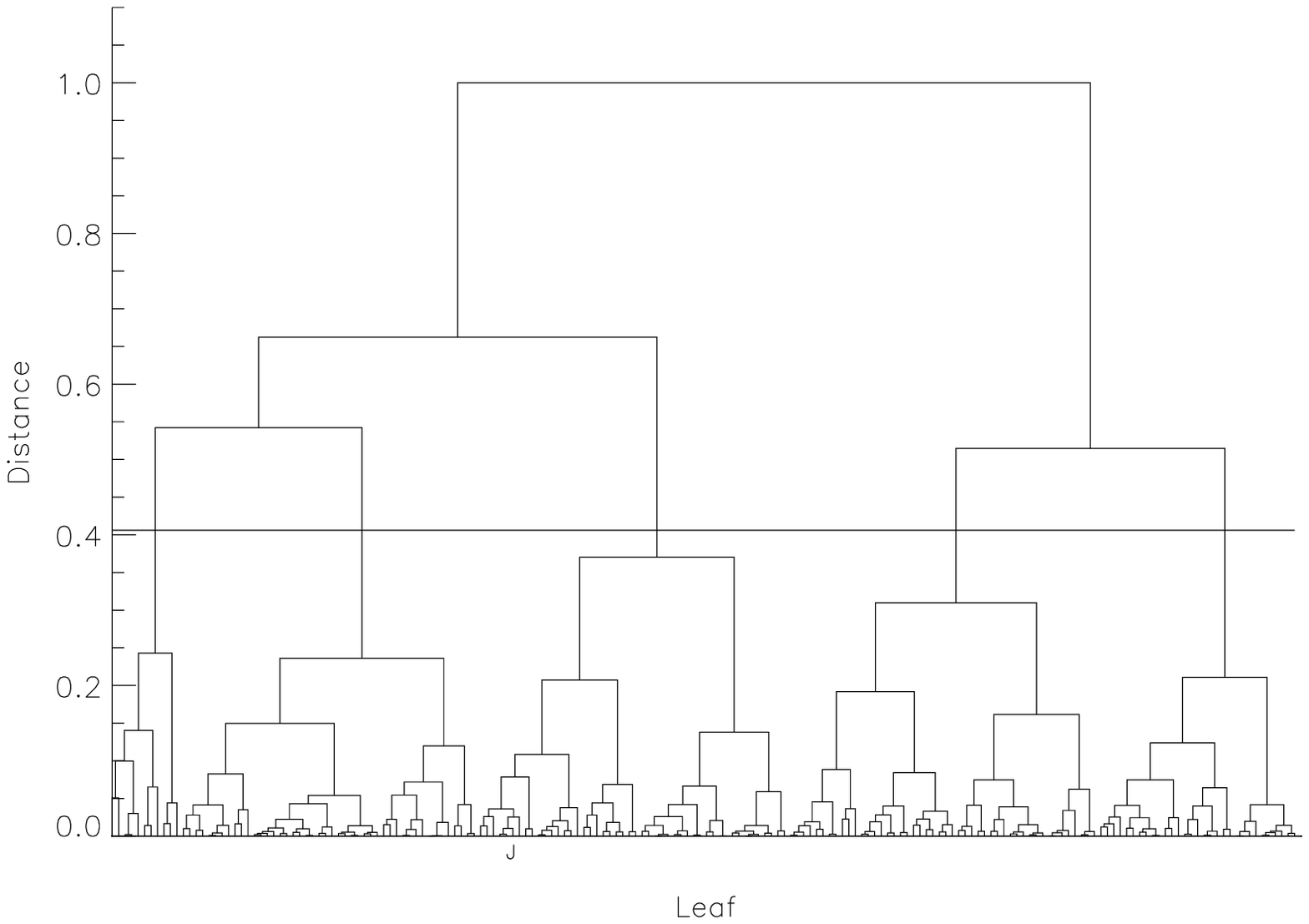}
\caption{Dendrogram of the best solution. Jupiter position  
is also reported. The horizontal line corresponds to the best
cut (see text). }
\label{dendro}
\end{figure}

First we decided  not to standardize the clustering  variables as this
may reduce the difference  among members, making the identification of
clusters  more  difficult.  Moreover,  as  a  general  rule, the  same
metrics should  be used for  the proximity matrix and  the inter-group
proximity measures.  We explored  different metrics and the effects of
different algorithms  of inter-group  merging.  The metrics  used are:
Euclidean,  City  Block,  Chebyshev,  and correlative  (based  on  the
Pearson   correlation   coefficient).    The  inter-group   algorithms
considered  are: single linkage,  complete linkage,  weighted pairwise
average and weighted centroid.  The full set of possibilities has been
investigated  by using  the cophenetic  coefficient and  analyzing the
corresponding dendrograms.  $\hat{c}$ ranged  from about 0.55 to 0.84.
Although  useful,  the  cophenetic  coefficient  cannot  be  the  only
parameter for finding the best solution; very different
hierarchies  may have  the same  $\hat{c}$. Here,  we  also considered 
other features in  order to define the best  solution.  Among them, we
first select  the dendrograms  having $N(h)$ distributions with the
lowest  variance,   i.e.  those more  peaked  for low  heights. Then,  we
required  the tails of the $N(h)$ distributions have to be very
``discrete'' for  large heights,  that is the  solutions with  $k$ and
$k+1$ clusters (for low values of  $k$) had to be well separated. Both
conditions help  to define a  robust solution which is  closely nested
for  small  heights and  stable  against  errors (e.g.   observational
errors) on the position of the EPs in the clustering space.\\
Taking into account these  constraints, the best solution was obtained
with  the Pearson  correlation distance\footnote{The  distance between
two  items  $i,j$  is  defined as  $d_{ij}=[(1-\phi)/2]^{1/2}$,  where
$\phi$ is the Pearson correlation coefficient.}  and weighted centroid
merging.
This solution corresponds to $\hat{c} = 0.83$.  According to the upper
tail rules,  the best cut produces five  robust clusters.  Remarkably,
the interval of  heights for which the solution  is stable corresponds
to  7\% of  the maximum  height  (notice that  the average  separation
between two  consecutive clusters is  of the order of  the maximum height
divided by $n$, or $\sim0.5\%$).
The  corresponding  best  solution   dendrogram  is  shown  in  figure
  \ref{dendro}.  The  dendrogram is  well structured: for  low heights
  the clusters are closely nested (i.e.  small variation of the height
  results in  a significant change  in the number of  clusters); while
  for higher values the clusters are well separated. \\
%
It  is  interesting  to   note  that  traditional  metrics  (like  the
Euclidean)  and  traditional  cluster  merging (like  single  linkage)
produce, in general, bad results.  It turns out that they are not able
to find structures for EPs. This  is in agreement with the analysis of
\cite{jia06}, although it was performed with a very different approach
to that adopted  here.  The main problem of  these hierarchies is that
either they produce just one or two big clusters and some outliers, or
the  solution is  so  nested that  the  definition of  cut-off is  too
arbitrary and meaningless.

\subsection{Testing the solution}

Before analyzing in  detail the best solution found,  we have to check
for its  robustness against the  absence of clusters.  We  recall that
cluster analysis will always produce a solution.
The  test  for  the  absence  of  structure is  done  by  an  accurate
evaluation of  the intra-cluster  separations.  In general,  unless we
are dealing with data with  very well separated clusters, the clusters
will tend  to have  some degree of  overlap.  The overlap  between two
clusters can  be estimated by  comparing the distance of  two clusters
with respect to their volume in the clustering space.  In practice, we
estimate  the center  of each  cluster ($\vec{c_i}$),  and  its radius
($r_i$), defined  as the sphere  which contains a given  percentage of
the cluster  members (both $\vec{c_i}$  and $r_i$ are computed  with a
Euclidean metric).   Thus the degree  of overlap between  clusters $i$
and $j$ is set  to: $ \omega_{ij} = (r_i+r_j)/ \|\vec{c_i}-\vec{c_j}\|
$.  Where for $ \omega_{ij}<1 $ the clusters $i,j$ are well separated.
The overall overlap of the solution is defined as:

$$ \omega = \sum_{i>j} \omega_{ij} \qquad  {\rm if} \quad  \omega_{ij} > 1. $$ 

\begin{figure*}[h]
\begin{center}
\includegraphics[scale=.70]{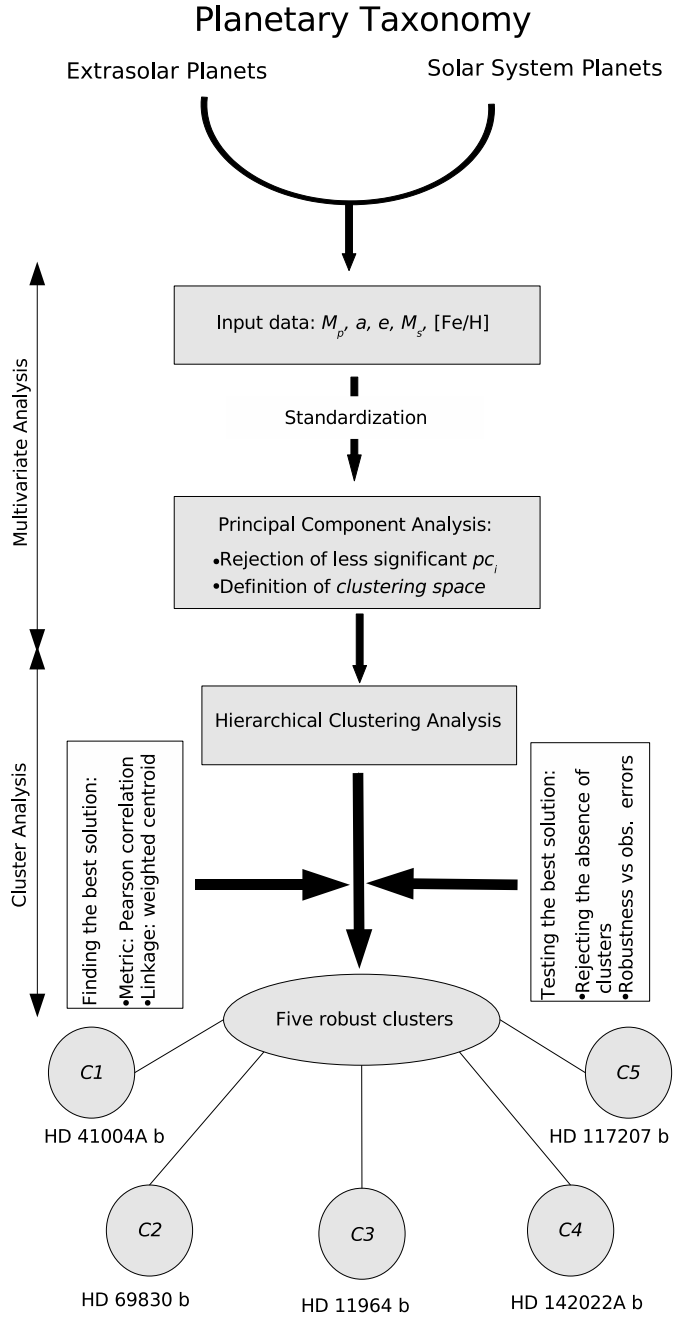}
\caption{Guidelines  for  the  planetary  taxonomy developed  in  this
  paper. As  for the Solar System,  only Jupiter has  been included in
  this  analysis so  far (see  text  for further  detail). Below  each
  cluster the correponding prototype is reported. }
\label{guide}
\end{center}
\end{figure*}

Higher values of $\omega$  imply larger overlap.  Notice that $\omega$
is a  cumulative parameter  and is sensitive  to the number  of single
intra-cluster  overlaps, as  it  should be.   To  clarify this  point,
imagine  a  solution  with  a  given degree  of  overlap  between  two
clusters, and  another one with the  same degree of  overlap but among
three clusters.  The  average degree of overlap will  be the same, but
the $\omega$  of the second solution  will be higher than  that of the
first.  Indeed the first solution is better than the second since only
two clusters are partially merged.  For the best solution, we obtained
$\tilde{\omega}=12.3$.  One way of  testing the absence of clusters is
with  Monte  Carlo  simulations.    We  then  performed  a  number  of
simulations (with $10^3$ solutions each), generating $n$ random points
in  the clustering  space.  The  points are  generated with  a uniform
distribution  along each axis.   Then we  run the  clustering analysis
using the same procedure described  before, varying also the volume of
the  Monte Carlo  generated points  in the  clustering space,  and the
percentage  of objects  used for  defining the  cluster's  radii.  The
probability of  obtaining $\omega <\tilde{\omega} $ is  less then 10\%
in  all cases.   Thus  we may  safely  reject the  possibility of  the
absence of structures.  For the shake of completeness, we also perform
additional  tests to  see  whether  our initial  choice  of using  $a$
instead   of   $P$   is   indeed   a   good   choice   in   terms   of
taxonomy. Performing clustering analysis  using $P$ instead of $a$, we
obtain a  solution which  has $\omega=20$ and  is compatible  with the
absence of structures.  A  possible explanation of this behavior could
be due to the  fact that $P$, $a$ and $M_s$ are  related. As a result,
the use  of the pair  $a,M_s$ is better  than the pair  $P,M_s$.  This
further strengthens the case for the use of the semi-major axis.\\
Moreover, we also tested the  solution with respect to the presence of
observational errors,  which can be very  large in some  cases.  To do
this, we  generated a random error  in the input  variables around the
nominal values for all the EPs.   We then repeated the PCA and cluster
analysis  of  these fictitius  samples  (one  for  each of  the  input
variables).   The solution obtained  has been  compared with  the best
solution.  For errors of a few  percent, we find that some planets may
move   to   other  clusters,   but   this   is   limited  to   a   few
objects. Increasing the errors, sometimes a cluster may split into two
sub-clusters.   For errors  larger than  10\% the  solution  may alter
considerably.  In general  the best solution does not  change much for
errors up to $\pm10\%$ for each  of the input variables.  This is very
important,  since it  means  that  the solution  is  quite stable,  in
particular considering  that we  are dealing with  projected planetary
masses and not real masses.\\
The  process   developed  for   EP  taxonomy  is   sketched  in
fig. \ref{guide}.\\

\section{Analysing the clusters}

\begin{figure}
\includegraphics[angle=-90,scale=.30]{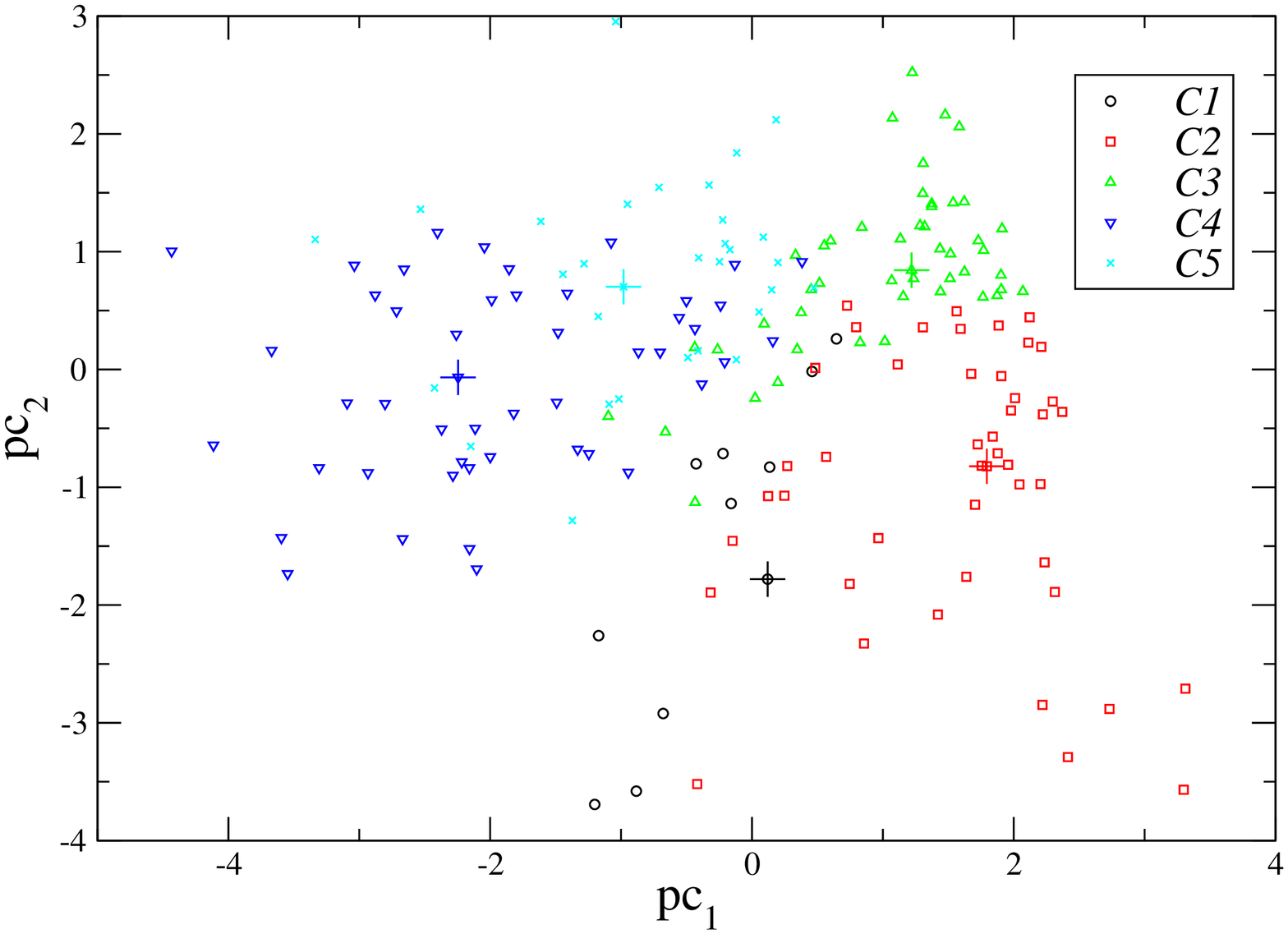}
\includegraphics[angle=-90,scale=.30]{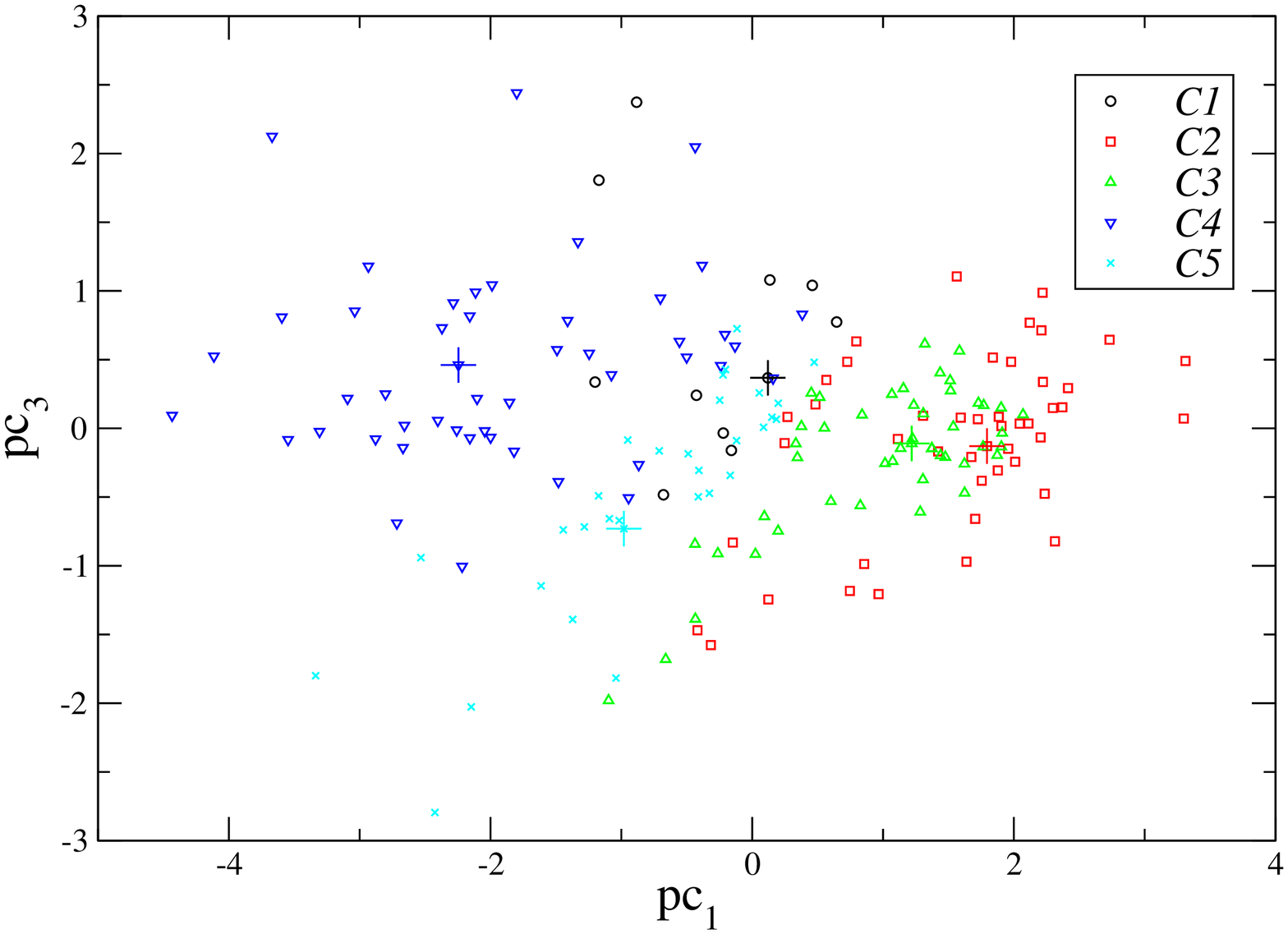}
\includegraphics[angle=-90,scale=.30]{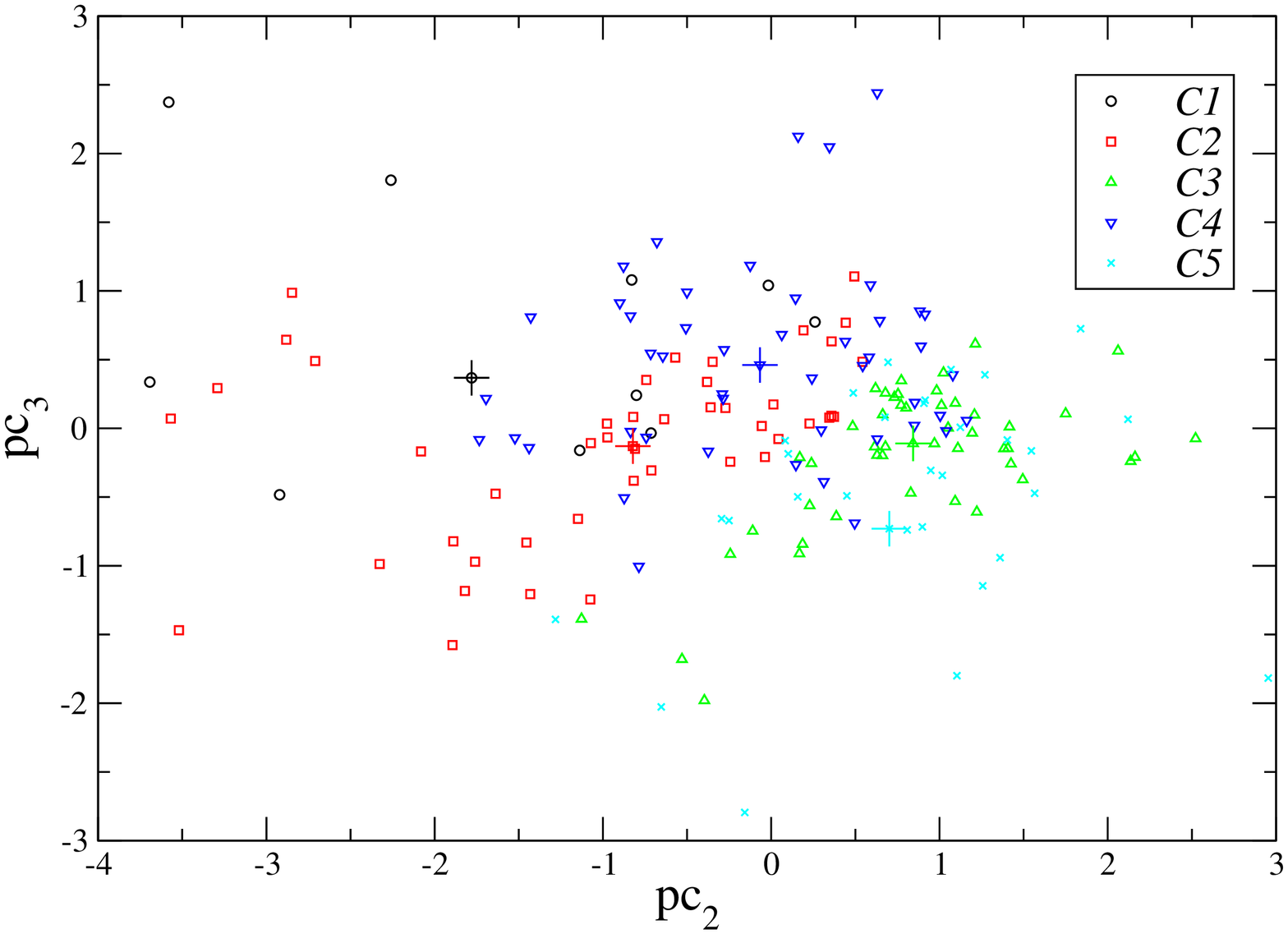}
\caption{Distribution of the clusters in the clustering space projected
onto the $pc_1-pc_2$, $pc_1-pc_3$ and $pc_2-pc_3$ planes. Large pluses
indicate the prototype planet of each cluster (see text).}
\label{pca_projected}
\end{figure}

Our best solution is composed of five robust clusters. In this section
we present the properties of each cluster.  Figure \ref{pca_projected}
shows the position of clusters in the clustering space.
%
We checked for inter-correlation among the input variables within each
cluster.  It is  commonly accepted that planets may  form in different
ways (core accretion vs disk instability), and that their evolution is
affected by  several parameters (disk density,  stellar types, opacity
etc).   The  EP database  may  reflect  such  complexity, however  the
signature of  these processes may  be blurred in  statistical analyses
which deal  with the  whole EP dataset.   On the contrary,  if cluster
separation  has something to  do with  the formation  and evolutionary
processes and is not just  a mere classification, it becomes important
to look  for trends within each  cluster.  In the  following we report
only highly  significant (with a  2-tailed probability less  than 5\%)
intra-cluster correlations.

\subsection{Cluster {$\cal C$}{\it 1}}

\begin{figure*}
\includegraphics[angle=-90,scale=.30]{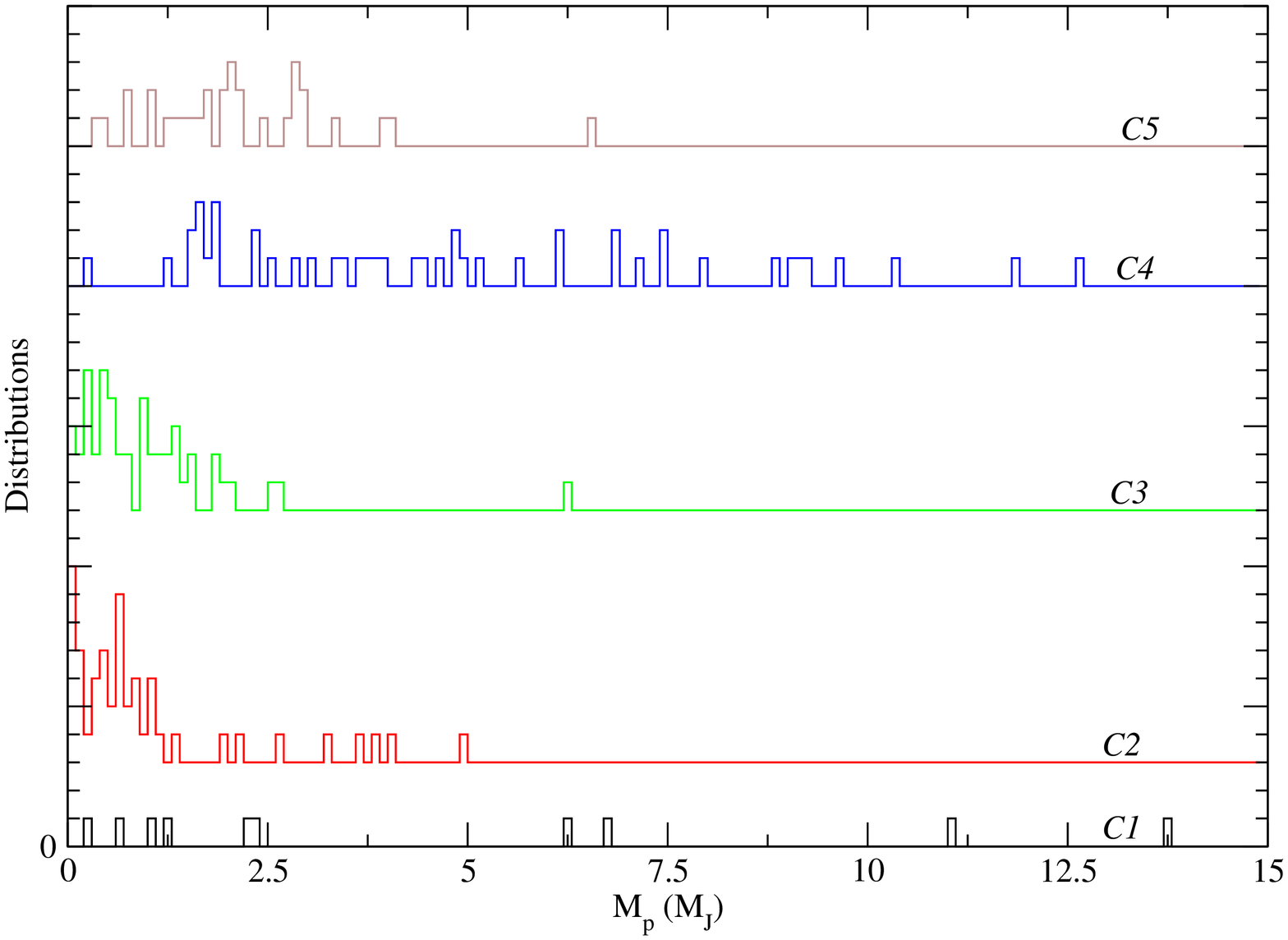}
\includegraphics[angle=-90,scale=.30]{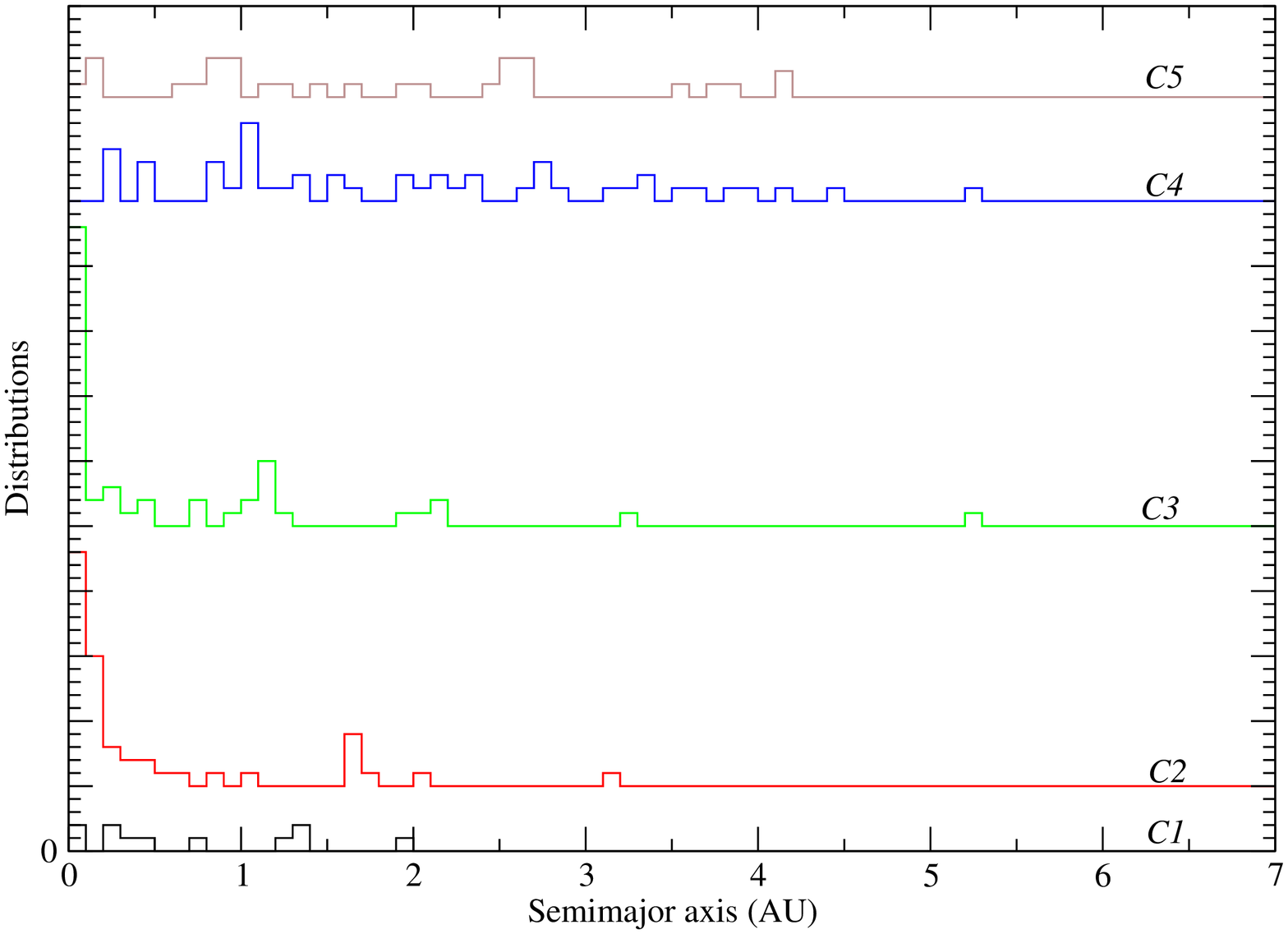}
\includegraphics[angle=-90,scale=.30]{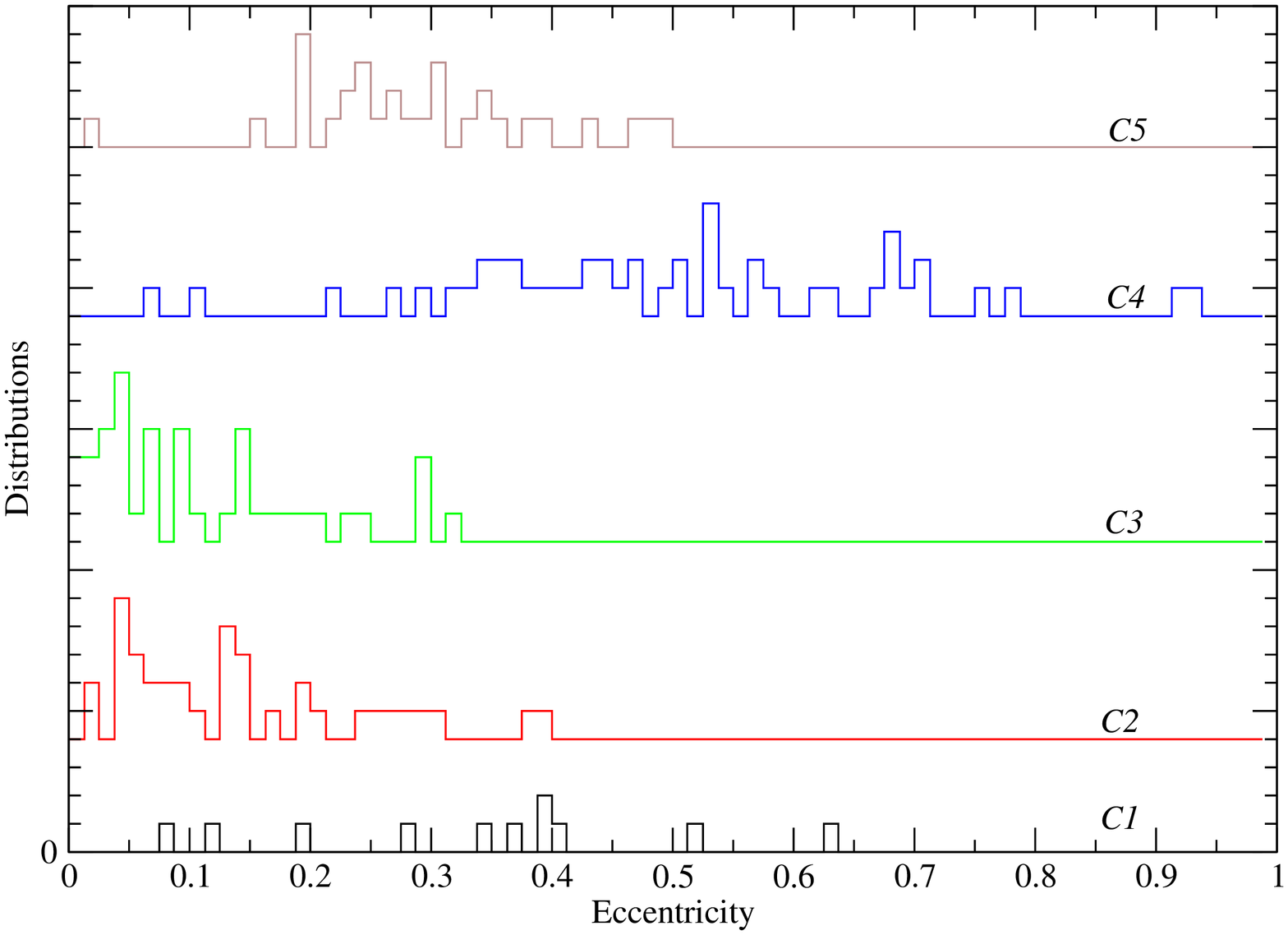}
\includegraphics[angle=-90,scale=.30]{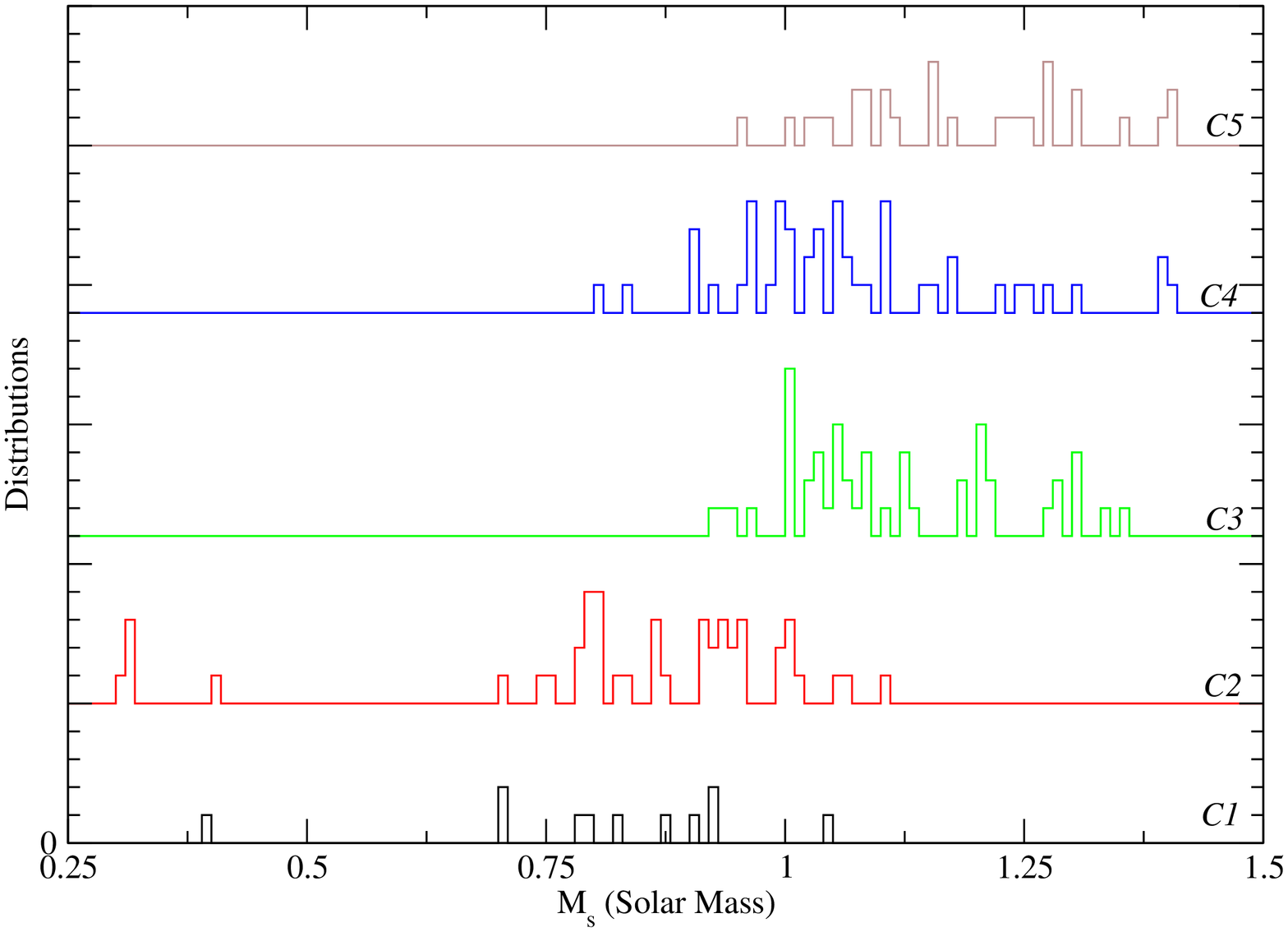}
\includegraphics[angle=-90,scale=.30]{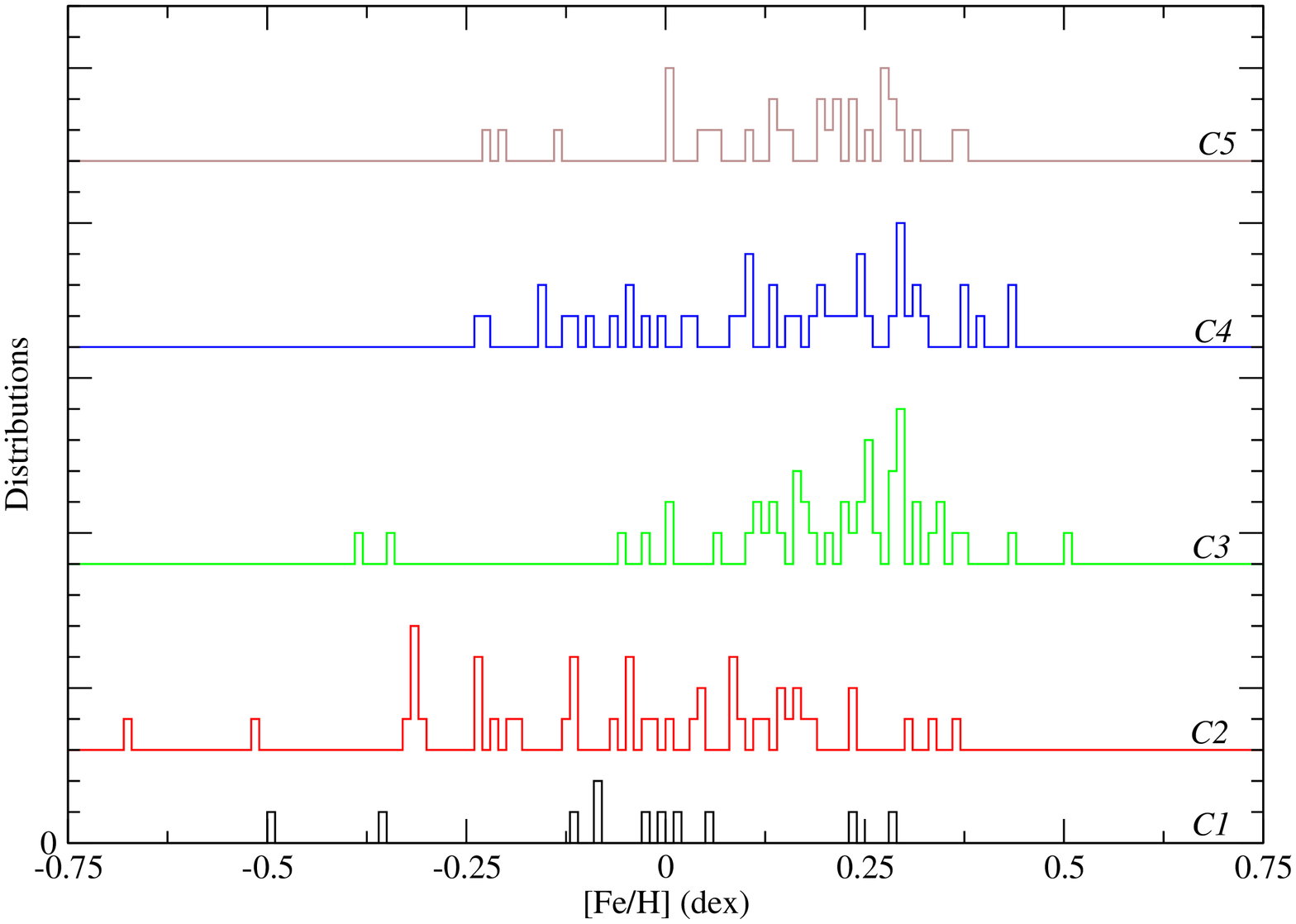}
\caption{Comparison of the distributions of the input variables for each cluster. 
Within each panel the distributions have been vertically shifted for
clarity. The vertical distance between two consecutive major ticks
corresponds to 5 units.}
\label{c_comparison}
\end{figure*}

\begin{figure*}
\includegraphics[angle=-90,scale=.35]{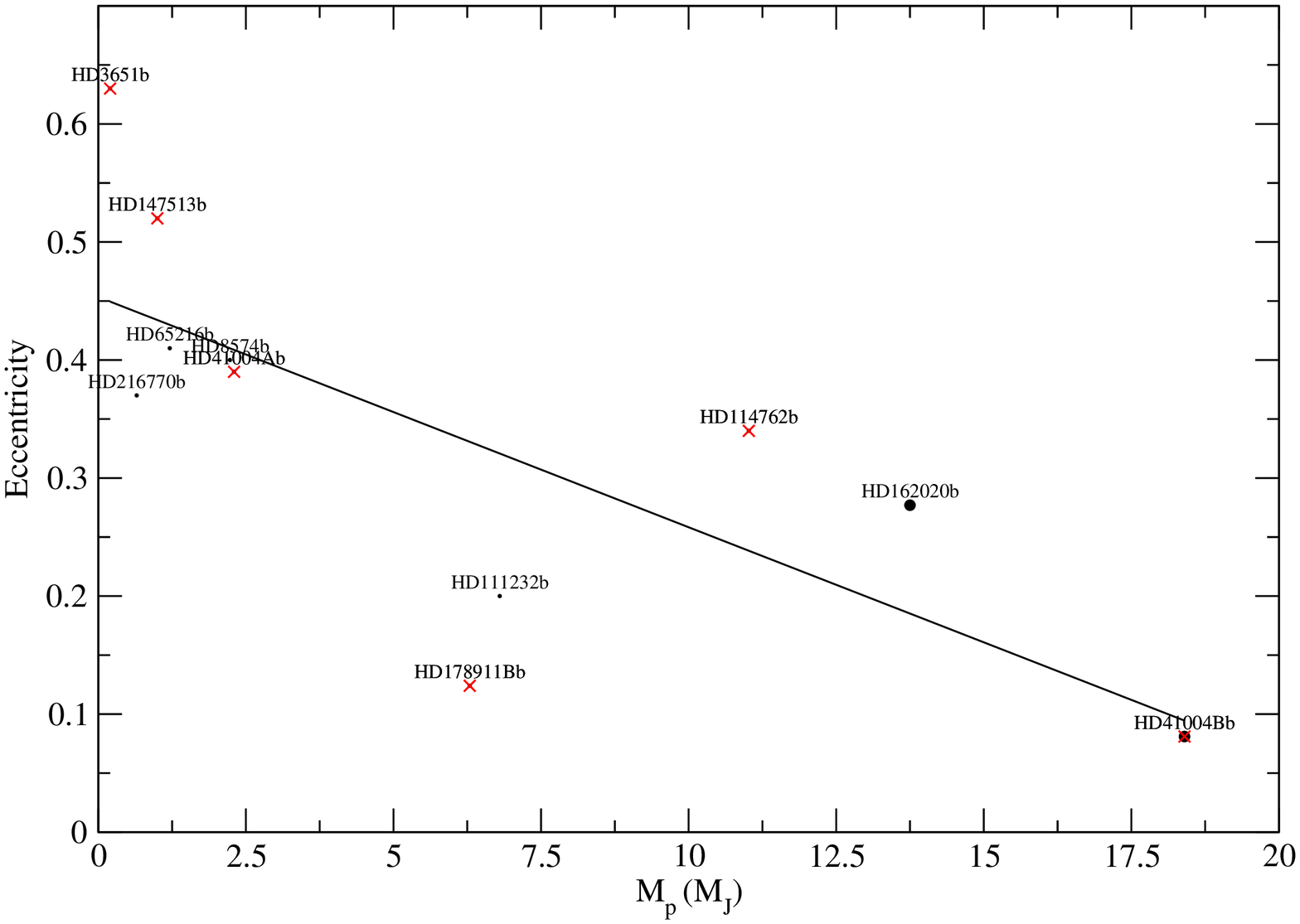}
\includegraphics[angle=-90,scale=.35]{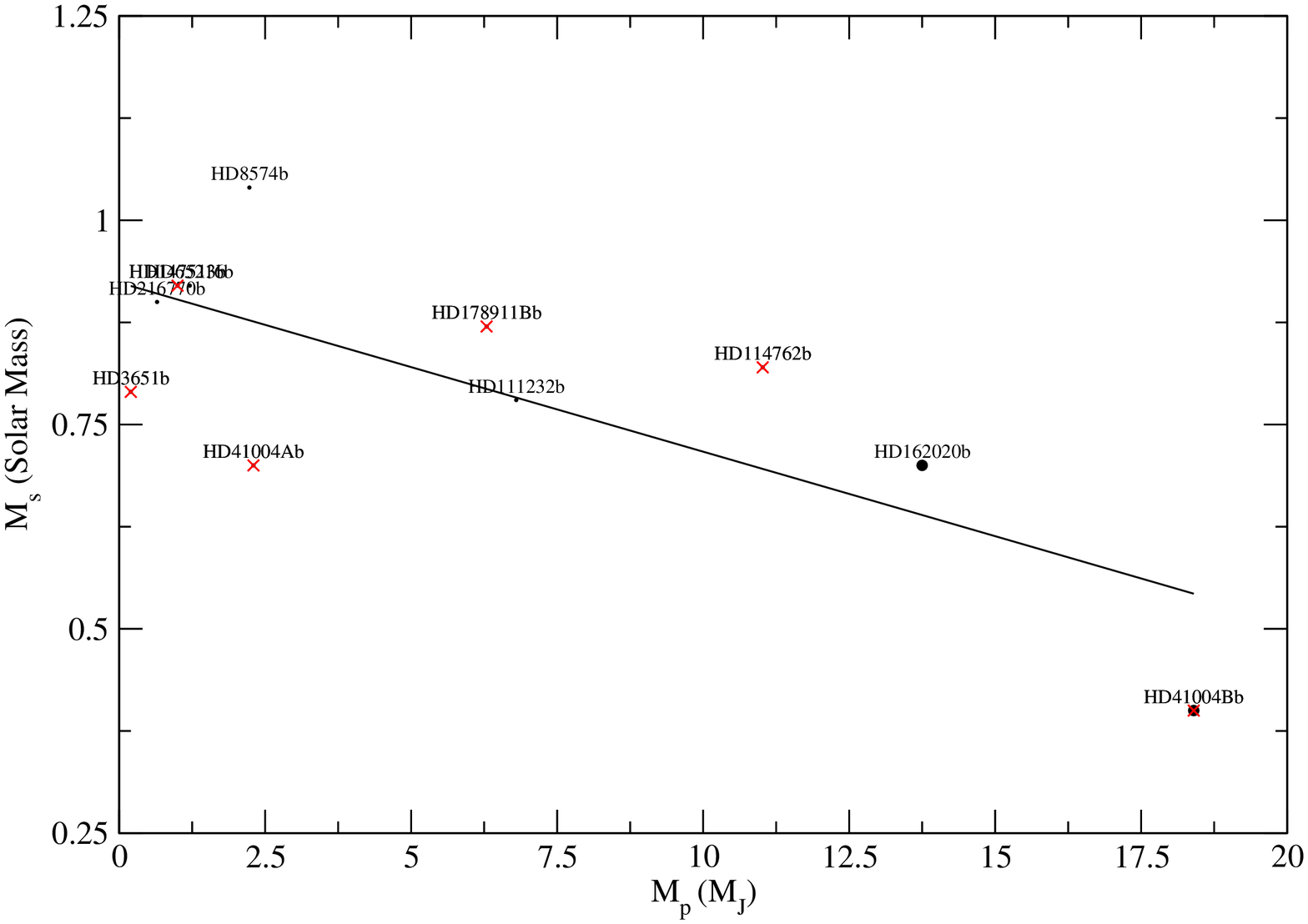}
\caption{Significant correlations within cluster {$\cal C$}{\it1}. The
  solid  line is  the  best  linear fit.  Black  circles indicate  hot
  Jupiters,  while  crosses correspond  to  planets  in multiple  star
  systems.}
\label{c1}
\end{figure*}

Containing  11  EPs (see  tables  \ref{tab}  and  \ref{tab_sup} for  a
detailed  description) this is  the least  populated cluster.   We can
define a  {\it prototype}  planet, that is  the object closest  to the
center  $\vec{c_i}$ of  the cluster.   The prototype  is HD~41004~A~b.
Figure \ref{c_comparison} shows the distribution of the input variables
for each cluster.  The EPs of this cluster are characterized by a wide
range  of  planetary  masses,  from  about  $0.2~M_J$  to  $18.4~M_J$.
The planetary semi-major axis ranges from 0.018~AU to 1.97~AU. The eccentricities
are quite spread, from 0.08 to 0.63. The stellar masses are remarkably
sub-solar (except  for HD~8574~b which  has $1.04~M_\odot$).  Finally,
star metallicity is very spread, from -0.5 to 0.28~dex.
{$\cal  C$}{\it  1}  contains  several  peculiar  EPs.   HD~41004~B~b,
HD~162020~b HD~114762~b are very  massive planets, with 18.4, 13.7 and
11~$M_J$ respectively  (see tabs.  \ref{tab}  and \ref{tab_sup}).  The
first two  are also hot Jupiters \citep[notice  that HD~162020~b could
be a brown dwarf, according to][]{udr02}.  This cluster contains 6 EPs
in multiple star systems (MSS), see tables \ref{tab} and \ref{tab_sup}
\citep[data on  multiple star systems have  been taken from][]{des06}.
Moreover HD~41004~B~b, HD~162020~b,  HD~114762~b and HD~111232~b orbit
low mass,  low metallicity, stars.  Despite this  variety, the objects
in  this  cluster  have  to  be considered  ``similar''  in  terms  of
clustering analysis  and ``different'' to the other  EPs. We underline
that  the cluster  analysis has  been  done in  the $\{pc_i\}$  space,
therefore  the   input  variables  may  vary   considerably  within  a
cluster.\\
 Significant  intracluster  correlations  exist  between  $M_p-e$  and
$M_p-M_s$ (see  fig.  \ref{c1}).  Notice that  $M_p$ is anticorrelated
with $M_s$.  This  is somehow unexpected as for  higher $M_s$ we would
expect  a higher  dust surface  density for  the  protoplanetary disks
\citep{ida05} and hence more  massive planets (consider also that here
we  have sub-solar  stellar  masses;  on this  point  see also  {$\cal
C$}{\it 4}).   We may  argue that  this has something  to do  with the
peculiar way  these EPs  formed, but no  firm conclusion can  be drawn
yet.

\subsection{Cluster {$\cal C$}{\it 2}}

\begin{figure*}
\includegraphics[angle=-90,scale=.35]{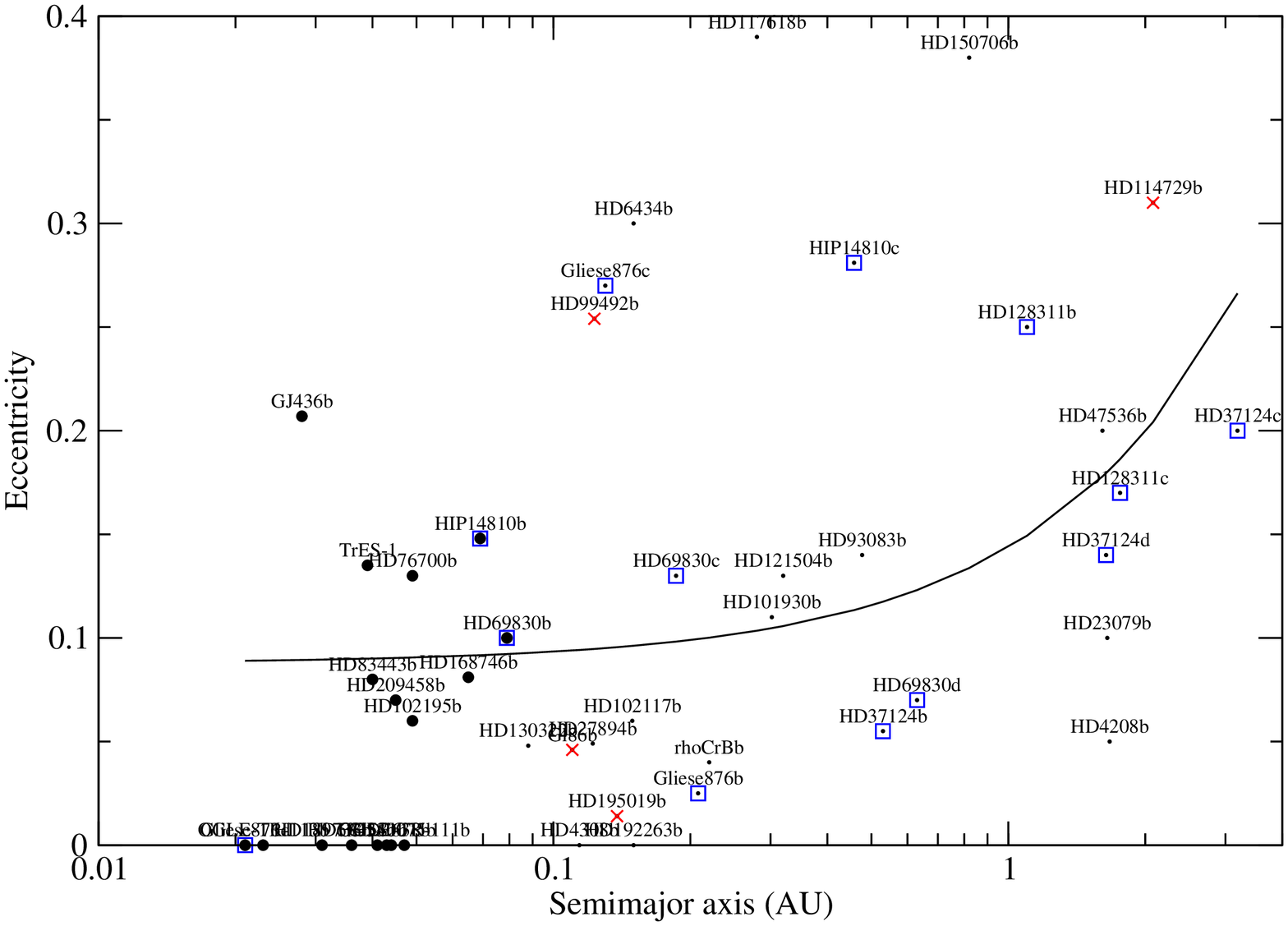}
\includegraphics[angle=-90,scale=.35]{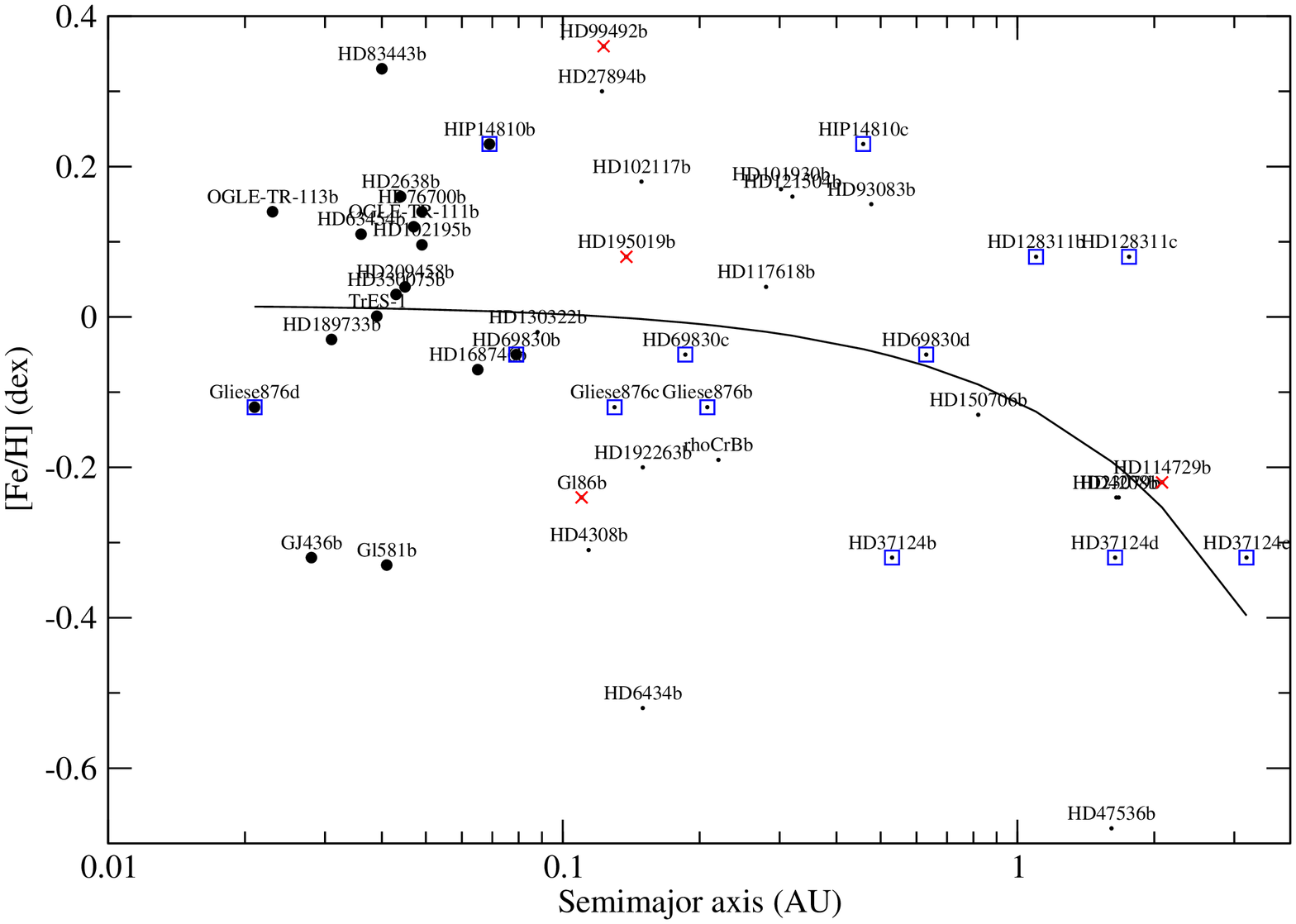}
\caption{Significant correlations within cluster {$\cal C$}{\it2}. The
solid  line  is the  best  linear  fit.   Black circles  indicate  hot
Jupiters, crosses correspond to MSS planets and squares to MPS planets
(notice the x-axis log scale).}
\label{c2}
\end{figure*}

This cluster contains 46 EPs (see tables \ref{tab} and \ref{tab_sup}).
The  prototype is  HD~69830~c.  This  cluster is  characterized  by an
$M_p$ distribution  with an  average of about  $1~M_J$ and  a standard
deviation  of $1.2~M_J$,  clearly peaked  at low  masses. Most  of the
planets have masses below $1.3~M_J$.  The semi-major axis distribution
is characterized by  two distinctive groups, one peaked  at very small
$a$ and the second, far less numerous, centered at $a\sim1.6$~AU.  The
overall  distribution  has  an  average  of  0.46~AU  and  a  standard
deviation  of 0.7~AU.   Planetary eccentricities  are moderate-to-low,
$0.11\pm 0.11$, and stellar  masses are remarkably sub-solar, $0.83\pm
0.2~M_\odot$.  Star metallicities are around solar, with an average of
-0.04~dex,   and  a   standard   deviation  of   0.22~dex  (see   fig.
\ref{c_comparison}).   Gl~581~b, Gliese~876~b-c-d  and  GJ~436~b which
orbit low mass stars  (respectively with 0.31, 0.32 and 0.41~$M_\odot$
they are  the lowest  $M_s$ in the  sample) also with  low metallicity
(respectively -0.33, -0.12, -0.32~dex) belong to this cluster.\\
{$\cal  C$}{\it 2}  contains  17 hot  Jupiters  (that is  37\% of  its
 members),  and  4  EPs  belong  to MSSs  (see  tables  \ref{tab}  and
 \ref{tab_sup}).  It  also contains 5 transiting  EPs \citep[the total
 number of  transiting EPs  is 14  -at December 2006-  but only  9 are
 involved in the present  analysis; see][]{bur06}.  These 5 transiting
 EPs do not seem to have  any particular properties, exept to have the
 highest stellar ages  of the sample (but we warn  that those ages may
 not  be well constrained).   Finally, {$\cal  C$}{\it 2}  contains 13
 planets in multiple planet systems (MPS).\\
The  significant intra-cluster correlations  are $a-e$  and $a-$[Fe/H]
 (see fig.  \ref{c2}).  The first  one is very interesting because $a$
 is  anti-correlated with  $e$.  Thus  planets further  away  from the
 stars  have  higher  eccentricities.   In  other  words,  either  the
 excitation of  $e$ is  more effective further  away from  the central
 star  (assuming that  planets  form in  circular  orbits) and/or  $e$
 dumping is more  effective for lower $a$.  This  result is consistent
 with tidal circularization of  planets with small $a$, however notice
 that at $a\sim0.02$~AU  the average $e$ is about  0.1, that is, still
 considerably non-zero \citep[we caution  that low $e$ values could be
 affected by  biases due to  fitting procedures,][]{for05}.  Moreover,
 with a  notably steep trend,  $a$ anti-correlates with  [Fe/H].  Thus
 either  the planetary migration  is more  pronounced for  high [Fe/H]
 \citep[which is in agreement with simulations, e.g.  see][]{liv03} or
 giant planets of  this cluster may form close  to stars ($\leq 1~$AU)
 in high metallicity environments.   We recall that for [Fe/H]$>0$ the
 most  accredited  formation theory  is  core accretion  \citep{fis05,
 san05}.  Notice also that the  distribution of the MPS planets in the
 $a-$[Fe/H] plane  is somehow opposite to the  overall observed trend:
 without  them the correlation  would be  even more  pronounced.  Both
 trends  are very  important because,  for the  first time,  they show
 significant  dependence between  metallicity  and orbital  parameters
 \citep[on the possible  existence of a period-metallicity correlation
 see][]{soz04}.  These findings also suggest that lower [Fe/H] planets
 tend  to  have  larger   orbits,  making  them  difficult  to  detect
 \citep[e.g. see][]{bos02}.  In turn,  this may affect the probability
 of  forming planets  with  respect to  the metallicity  \citep{san04,
 fis05}.

\subsection{Cluster {$\cal C$}{\it 3}}

\begin{figure*}
\includegraphics[angle=-90,scale=.35]{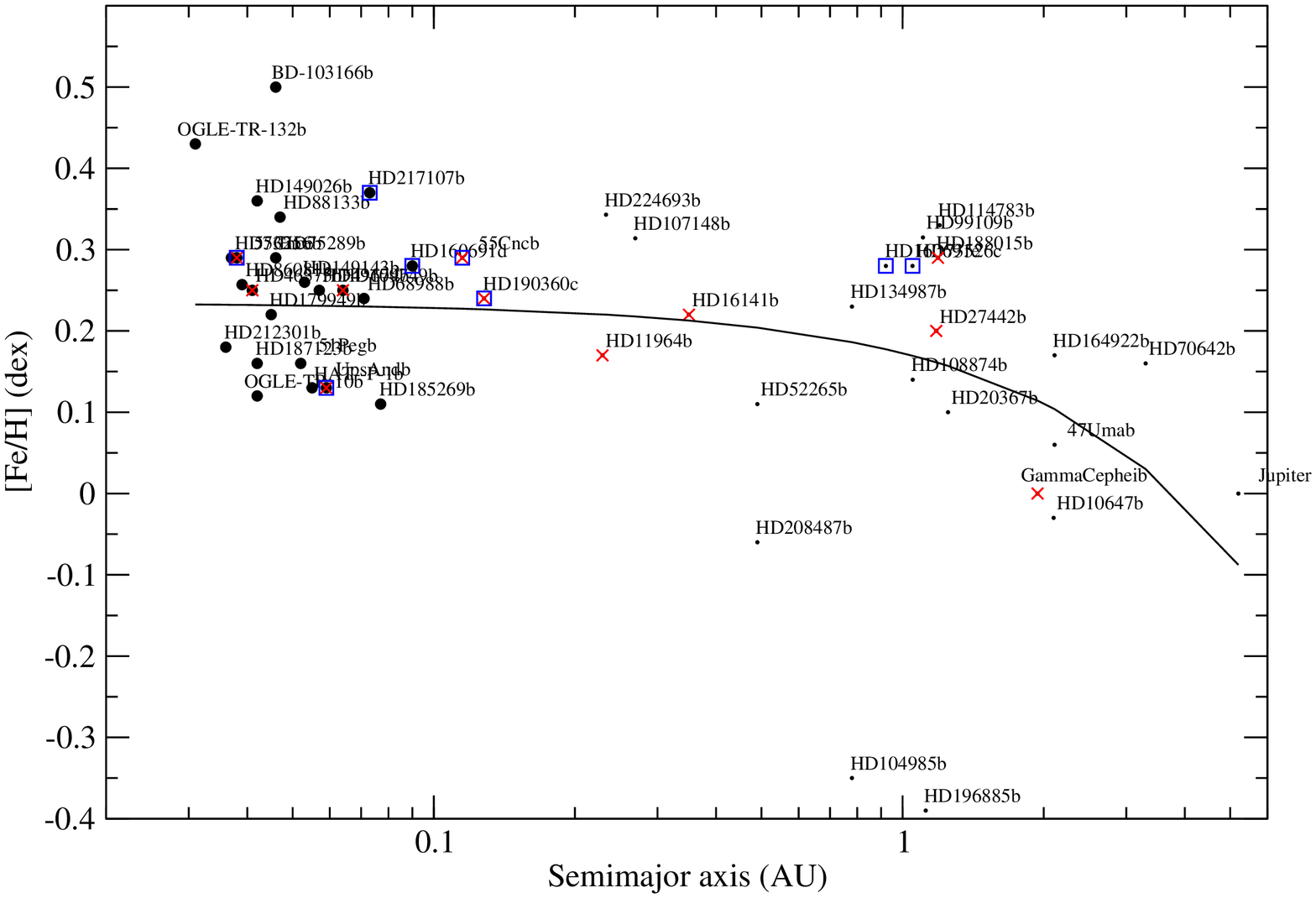}
\includegraphics[angle=-90,scale=.35]{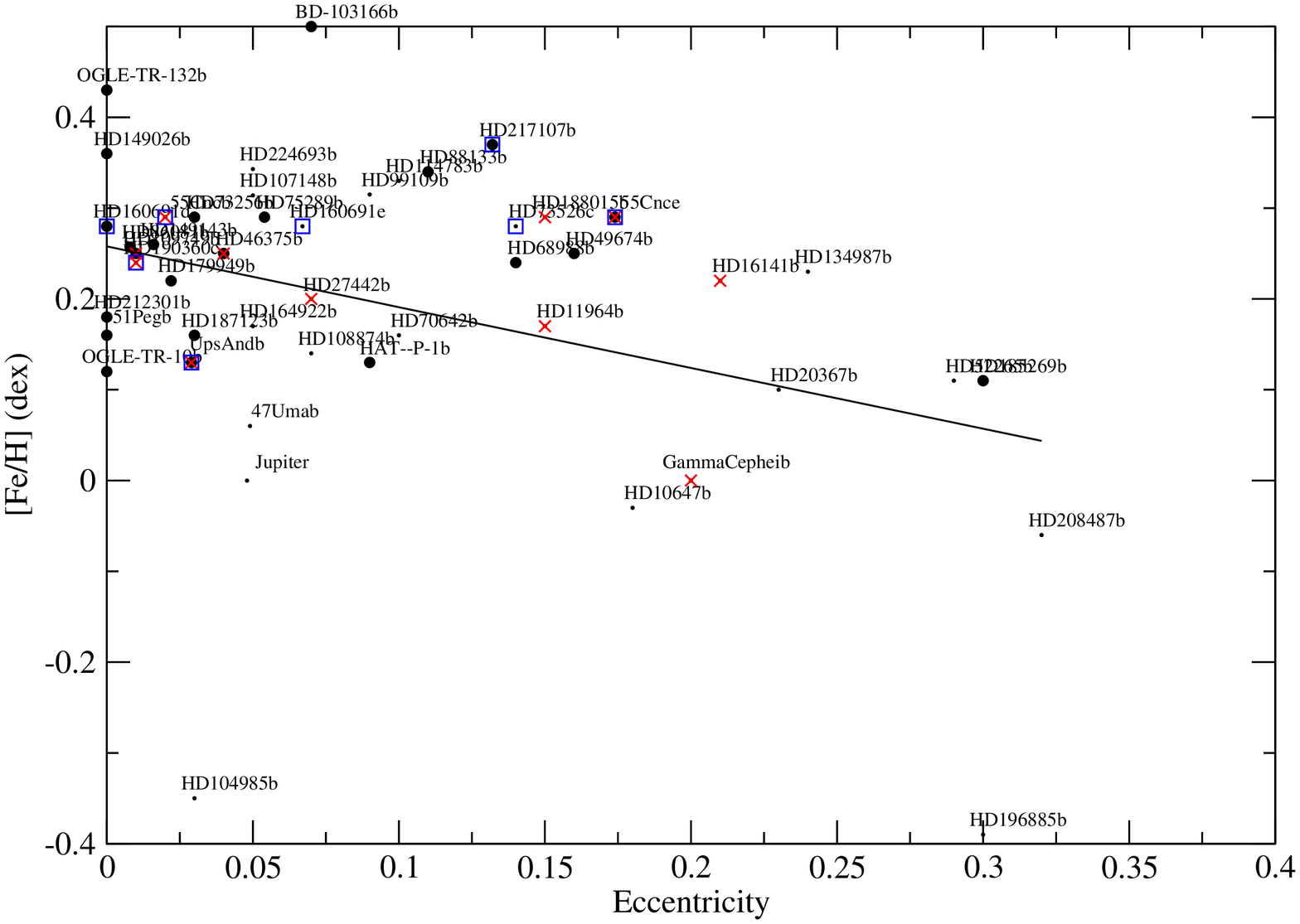}
\includegraphics[angle=-90,scale=.35]{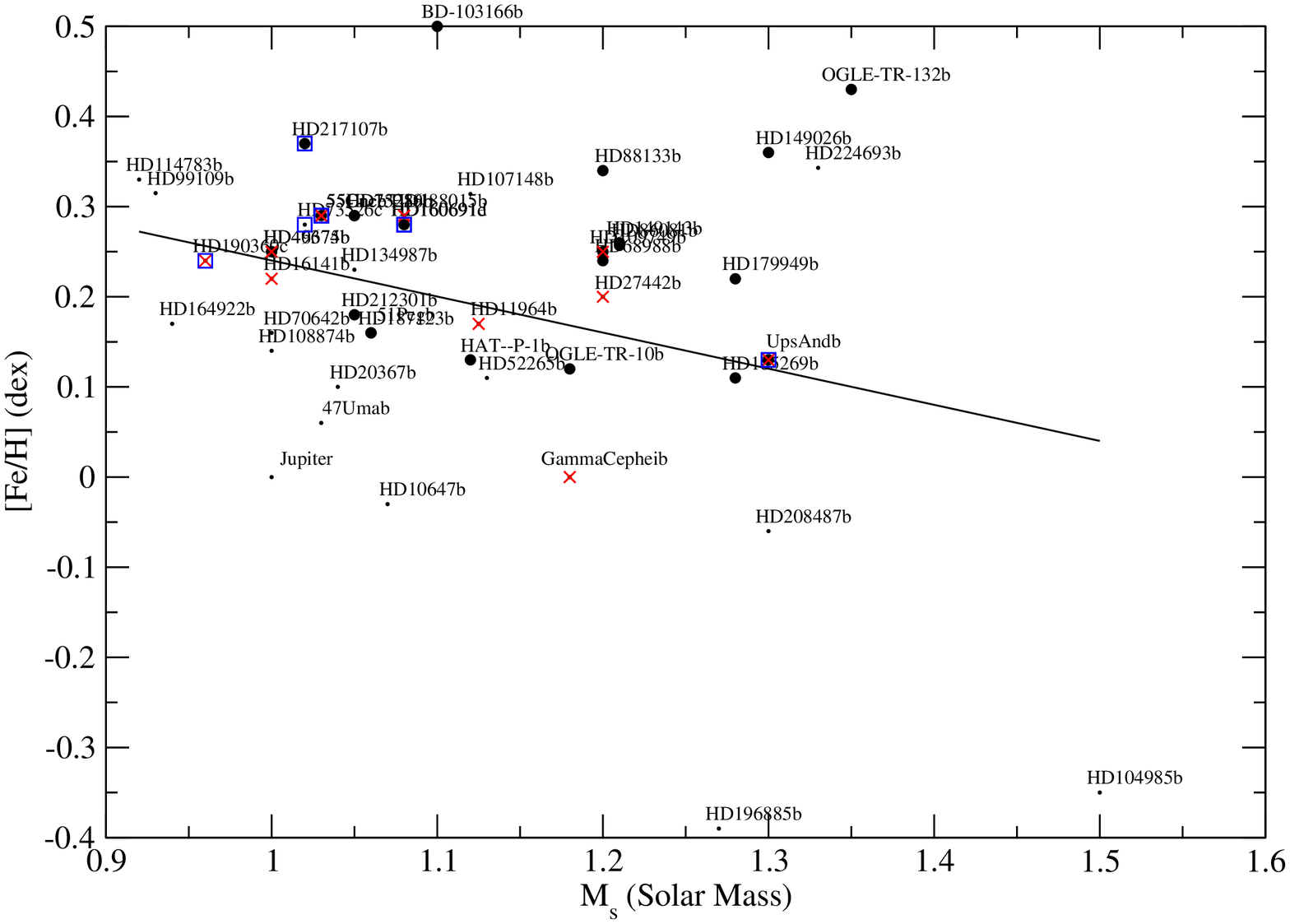}
\caption{Significant correlations within cluster {$\cal C$}{\it3}. The
  solid  line is  the best  linear  fit.  Black  circles indicate  hot
  Jupiters,  crosses correspond  to  MSS planets  and  squares to  MPS
  planets (notice the x-axis log scale of the $a-$[Fe/H] plot).}
\label{c3}
\end{figure*}

Containing  48  EPs  (see  tables \ref{tab}  and  \ref{tab_sup})  this
cluster, along  with {$\cal  C$}{\it 4}, is  the most  populated.  The
prototype is  HD~11964~A~b, and it contains Jupiter.   This cluster is
characterized by an $M_p$ distribution peaked at low masses: basically
all EPs are below $2.5~M_J$.  The semi-major axis distribution is very
peaked at  low values: most  of the bodies  are below 0.25~AU,  with a
second, less  numerous, peak at about 1~AU.   Eccentricities are below
about  0.3, having an  average of  0.096 and  a standard  deviation of
0.091.  Stellar   masses    span   from   0.98    to   $1.24~M_\odot$.
Metallicities tend  to be super-solar  and vary from 0.02  to 0.35~dex
(see fig \ref{c_comparison}).  \\
The    significant   intra-cluster   correlations    are:   $M_p-M_s$,
$M_p-$[Fe/H],  $a-M_s$,   $a-$[Fe/H],  $e-$[Fe/H],  and  $M_s-$[Fe/H].
However  the  first  three  are  due to  outliers:  without  them  the
correlations  do   not  hold.   Therefore  they  are   not  considered
interesting.   On the contrary,  the latter  three -all  involving the
metallicity- are robust (see fig. \ref{c3}).
First of  all, $a$  is anti-correlated with  [Fe/H].  Thus  either the
planetary  migration is  more  pronounced for  high  [Fe/H] (see  also
{$\cal  C$}{\it  2}) or  planets  with  lower  [Fe/H] form  at  larger
distances. A  striking result is that  Jupiter fits very  well in this
cluster:  its large  $a$ would  be the  result of  its formation  in a
solar-like  metallicity disk.  Moreover,  $e$ is  anti-correlated with
[Fe/H].   In other  words,  higher metallicities  correspond to  lower
eccentricities.  Thus  either the excitation of $e$  is more effective
-or dumping is  less effective- for lower metallicities.   In order to
understand the meaning of  these correlations, two further points have
to be  considered.  First, the existence of  $a-$[Fe/H] and $e-$[Fe/H]
correlations does not imply a correlation between $a$ and $e$.  Indeed
they  are not  correlated to  the  level of  confidence adopted  here.
Moreover,  the   members  of   this  cluster  have   very  super-solar
metallicities.   These  correlations,  if  related  to  the  formation
processes,  may give important  indications about  the origin  of high
eccentricities. Many theories  have been proposed \citep{hol97, mur02,
gol03, zak04,  nam05, ada06, dan06, fre06} but  none has observational
support yet. Here we suggest that  the metallicity acts in some way in
determining $e$.   For instance, since  high [Fe/H] produces  a faster
migration, the low  values of $e$ observed for high  [Fe/H] may be the
result   of  the  migration   process,  e.g.    tidal  circularitazion
\citep[see also][]{hal05}. Alternatively the condition for the pumping
up of $e$ during  planet-disk interactions \citep{sar04, mat06} is not
achieved  in high  [Fe/H] environments:  indeed higher  [Fe/H] implies
higher disk viscosity and hence  a lower probability of $e$ excitation
\citep[see  equ. 8  and  9 of][]{sar04}.  If  confirmed, these  trends
suggest  that planetary  orbital parameters  are mainly  controlled by
disk properties (e.g.  metallicity) rather than being affected by {\it
external}  factors,  like  perturbation  due to  interactions  with  a
companion star or star encounters.\\
Finally,  $M_s$ is  anti-correlated with  [Fe/H].  It  is not  easy to
understand this correlation, and in  particular if it has something to
do with  the planetary  formation process.  We  note that EPs  in this
cluster (which  are quite  confined in terms  of $a$ and  $M_p$) orbit
stars whose metallicities tend to increase as stellar masses decrease.
If we  assume that higher  $M_s$ implies higher  proto-planetary disks
masses from  which EPs  formed, this implies  that to form  planets in
lower  metallicity environments a  more massive  disk is  required, in
agreement with the core accretion theory.\\
Eleven EPs are in MSSs  (see table \ref{tab_sup}). It contains also 23
hot Jupiters (48\%  of the members) and 8 MPS  planets.  They are well
spread and seem not to affect the $a-$[Fe/H], $e-$[Fe/H] correlations.
However, all  MPS planets and most  of the hot Jupiters  lie above the
linear fit in  the $M_s-$[Fe/H] plane. Notice also  the MPS planets of
this cluster  have $e<0.2$ and  [Fe/H]$>0.2$ and for this  reason they
differ  from   those  of  {$\cal  C$}{\it  2}   (which  have  $e<0.3$,
[Fe/H]$<0.2$) and those of {$\cal C$}{\it 4} (which have $e>0.3$).

\subsection{Cluster {$\cal C$}{\it 4}}

\begin{figure*}
\includegraphics[angle=-90,scale=.35]{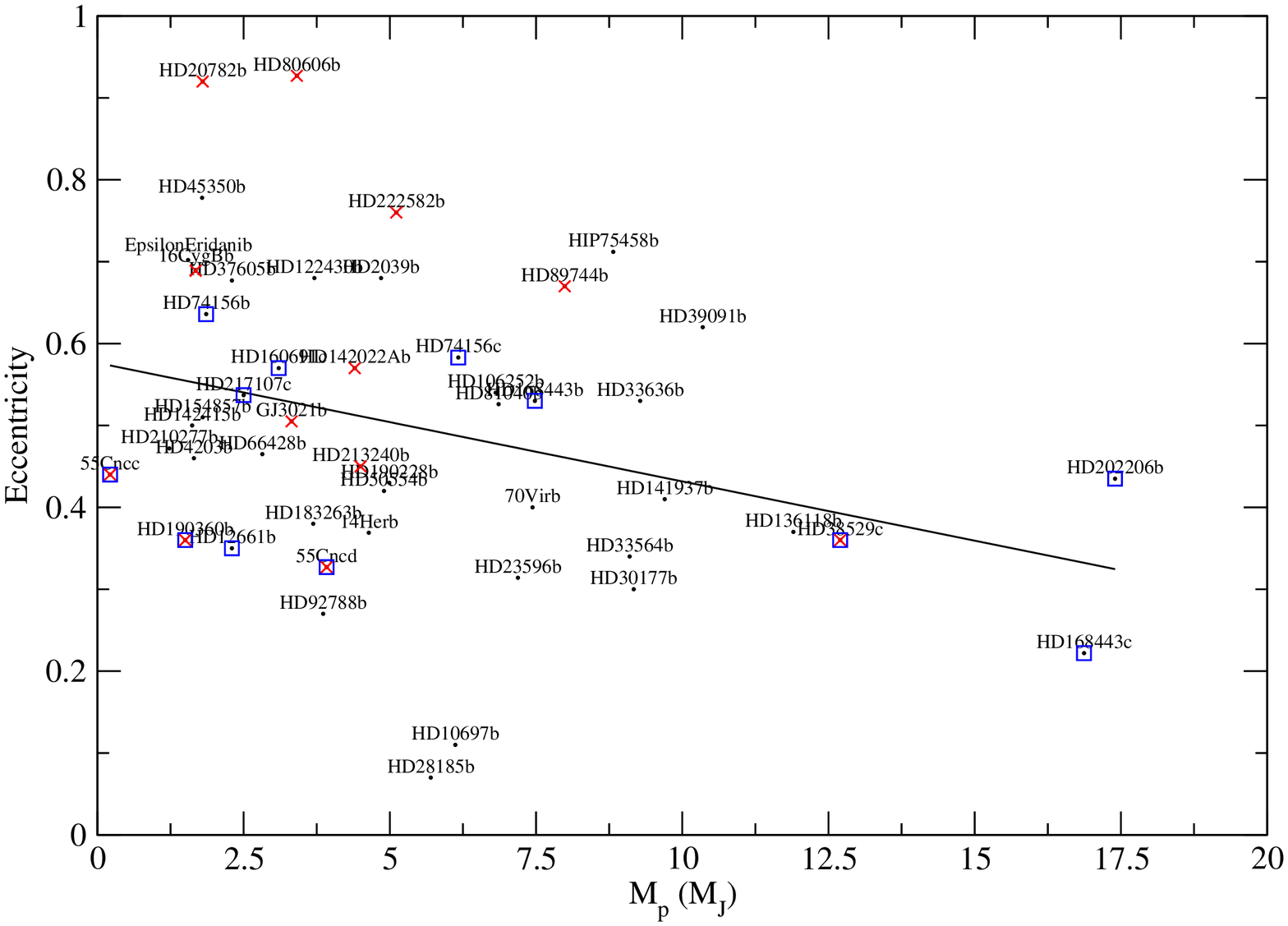}
\includegraphics[angle=-90,scale=.35]{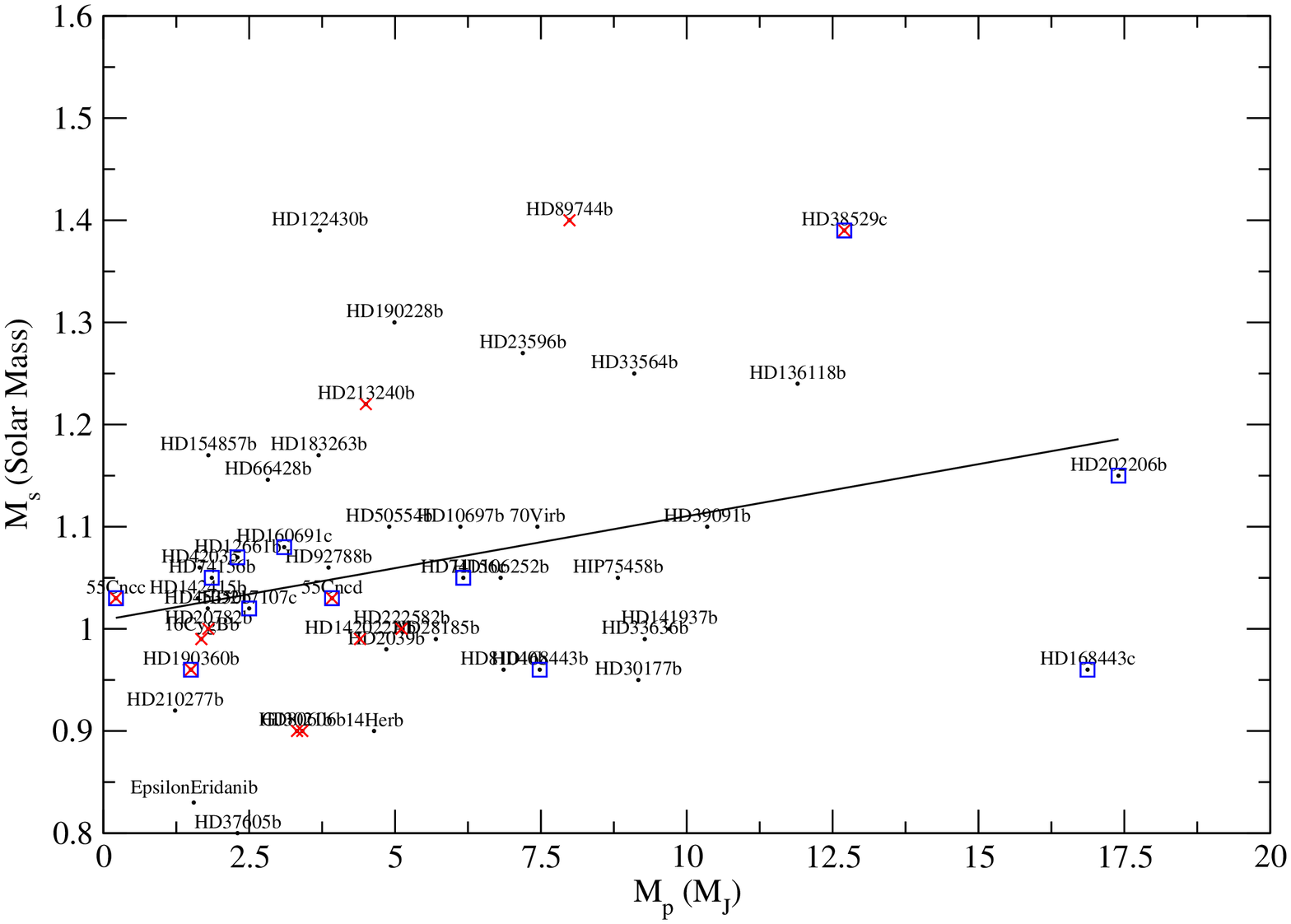}
\caption{Significant  correlations  within  cluster {$\cal  C$}{\it4}.
The  solid line is  the best  linear fit.   Crosses correspond  to MSS
planets and squares to MPS planets. }
\label{c4}
\end{figure*}

This cluster contains 48 EPs (see tables \ref{tab} and \ref{tab_sup}).
The prototype is HD~142022~A~b.  {$\cal C$}{\it 4} is characterized by
rather  flat distributions  of the  input  variables (see
fig. \ref{c_comparison}).   In terms of planetary  masses, it contains
very  massive bodies,  basically all  having $M_p>2.0~M_J$.  Apart from
very few  exceptions, it  contains all the  EPs of the  dataset having
mass greater than $5~M_J$. The average and standard deviations are 5.45
and  $3.92~M_J$, respectively.   The semi-major  axis  distribution is
also quite  flat, ranging from about  0.5 to 5~AU.   Mean and standard
deviations  are  1.98 and  1.27~AU,  respectively.  Eccentricities  are
remarkably  moderate-to-high,  spanning  from  0.3 to  0.8  (mean  and
standard deviation  are 0.49 and 0.18,  respectively). Stellar masses
are mainly  around one  solar mass, with  a slight  overabundance of
super-solar mass  objects.  Mean and  standard deviations are  1.06 and
$0.13~M_\odot$,  respectively.   Metallicities   span  from  -0.25  to
0.3~dex,  having  an  average  and  standard  deviations  of  0.14  and
0.17~dex,  respectively.  Thus  despite   its  wide distribution  of
metallicities, EPs are mainly super-solar.\\
Two  significant  correlations exist  for  this  cluster: $M_p-e$  and
$M_p-M_s$ (see  fig. \ref{c4}). The  first implies that  lower mass
EPs have  higher $e$,  thus the mechanisms  for the pumping-up  of the
eccentricity are  more active  in low mass  planets, at least  for the
high  semi-major  axes and  moderate  positive  metallicities of  this
cluster. Moreover, EP masses are correlated with stellar masses. This
may confirm the fact  that higher $M_s$ implies larger proto-planetary
disk surface density and hence larger $M_p$ \citep{ida05}.\\
{$\cal C$}{\it  4} contains 12 EPs in  MSSs and 12 in  MPS (see tables
\ref{tab}  and \ref{tab_sup}).   Notice that  the MSS  planets  may be
responsible of the $M_p-e$ correlation  as many of them have low $M_p$
and high $e$.

\subsection{Cluster {$\cal C$}{\it 5}}

This cluster contains 31 EPs (see tables \ref{tab} and \ref{tab_sup}),
and the  prototype  is HD~117207~b.   Planetary  masses have  intermediate
values,  with  a  mean  of  $2.16~M_J$ and  a  standard  deviation  of
$1.24~M_J$, respectively.  The  semi-major axis distribution is rather
flat  spanning  from  0.37  to  3~AU, with a few  bodies  around  4~AU.
Eccentricities  are peaked  at  0.2-0.3,  and range  from  0.2 to  0.5.
Stellar masses  are super-solar, having mean  and standard
deviation   of  1.22   and   $0.21~M_\odot$,  respectively.    Stellar
metallicities  are  also   remarkably  super-solar,  having  mean  and
standard     deviation      of     0.15     and      0.15~dex     (see
fig. \ref{c_comparison}).\\
The formally significant  correlations are: $M_p-M_s$, $M_p-$[Fe/H] and
$a-e$. However they are all due  to outlier planets and thus cannot be
considered as  real (notice that if  we do not  consider the outliers,
the correlation  $a-e$ is  very close to  the 5\%  significancy level,
thus it  may become significant as   more objects are added to
this cluster by future discoveries).\\
It  contains  7  EPs  in  MSSs  and  only  one  hot  Jupiter  (namely,
HD~118203~b).  It is interesting to understand why this hot Jupiter is
in this  cluster.  The peculiarity of  HD~118203~b is that  it has the
highest  eccentricity (0.309)  among hot  Jupiters.  Notice  that 
HD~185269~b (in  {$\cal C$}{\it  3}) also has a  very high  eccentricity of
0.3.  All  the other  input variables  are  the same,  except for  the
planetary mass.  HD~118203~b  with a mass of $2.13~M_p$  is one of the
most massive hot Jupiters.  This  explains why HD~118203~b has been put
in this cluster. This cluster contains 13 EPs in MPS.\\
Despite  the lack of  any correlation,  few comments  may be  added to
uncover  the nature  of this  cluster.  {$\cal  C$}{\it 4}  and {$\cal
C$}{\it 5} have some similar  traits. They have spread and rather flat
distributions   of  input  variables.    Moreover,  both   have  large
semi-major  axes  and  eccentricities.   However,  {$\cal  C$}{\it  5}
contains objects with higher  stellar masses, lower eccentricities and
lower masses than {$\cal C$}{\it 4}.

\section{Discussion and conclusion}

In this paper we develop  the basis for an extrasolar planet taxonomy.
We use as many inputs as possible for this analysis, in particular the
planetary  mass,  semi-major  axis,  eccentricity,  stellar  mass  and
stellar  metallicity. We  identify  the best  procedure  to follow:  a
multivariate  statistical analysis  (PCA) to  find the  most important
variables,  and then  hierarchical cluster  analysis.  We  analyze the
solutions via canonical means (through the cophenetic coefficient) and
by analysis of the distribution of the dendrogram's heights.  The best
result is achieved with non traditional metric and merging algorithms,
namely the  Pearson correlation  metric and weighted  centroid cluster
merging.   We reject the  absence of  clustering structure  with Monte
Carlo simulations, and also test the stability of the solution against
observational errors of the input variables. The procedure we followed
is able  to provide a roboust  extrasolar planet taxonomy  even if the
number of  planets is  still low. The  general traits of  the taxonomy
developed here will be updated as more planets become available. \\
Our  best  solution  consists  of  five clusters.   We  discuss  their
properties with  respect to  the physically relevant  input variables.
We show the importance of including the environmental variables ($M_s$
and  [Fe/H]) to  discriminate between  otherwise similar  planets; and
also to merge together different  bodies (like EP in MSSs and orbiting
single stars).  For instance, we  were able to divide the hot Jupiters
into -at least- 2 main  groups (see tab. \ref{tab}).  This division is
mainly due  to the stellar  mass and metallicity.  Those  belonging to
{$\cal C$}{\it 3} basically orbit around stars with super-solar masses
and  high  metallicities; those  of  {$\cal  C$}{\it  2} orbit  mostly
sub-solar  mass  stars  with  moderate (both  positive  and  negative)
metallicities. This may reflect differences in the formation processes
of these EPs.\\
Jupiter belongs to cluster {$\cal  C$}{\it 3}. Much as been speculated
about  the  similarity of  our  Solar  System  and extrasolar  systems
\citep[e.g.][]{bee04},   in   particular   concerning  the   formation
histories. With the help of cluster analysis we may identify those EPs
which are more similar -in the 3-fold clustering space- to Jupiter. We
suggest that the actual large semi-major axis of Jupiter is the result
of its formation in a solar-like metallicity disk.\\
We  also  analyzed  the  intra-cluster correlations,  since  this  may
provide  important information  about the  formation and  evolution of
bodies  within  a  cluster.   This  is crucial  in  order  to  uncover
information which may be  hidden in the ``blind'' statistical analysis
performed on  the whole EP  database. The most  important correlations
found are  those for {$\cal C$}{\it  2}, {$\cal C$}{\it  3} and {$\cal
C$}{\it 4}  (see tab. \ref{tab}).   Remarkably, for {$\cal  C$}{\it 2}
and {$\cal C$}{\it 3} we find important trends between metallicity and
orbital parameters.  We find that [Fe/H] has very important effects on
the  semi-major axis  (and thus  on the  migration processes)  and the
eccentricity.  It may also happen that the same variables correlate in
an  opposite way  between two  different clusters  (see  the $M_p-M_s$
correlations for {$\cal C$}{\it  1} and {$\cal C$}{\it 4}).  Moreover,
we also studied  the distribution of planets in  multiple star systems
in each  cluster. They do  not seem to  play a particular role  in the
corresponding cluster correlations.  Similar considerations apply also
for multiple planet systems.\\
In addition  to these main five  clusters, we may see  the position of
the pulsar  planets in the  clustering space. Obviously  these planets
were  not included in  the previous  analysis because  we do  not have
$M_s$ and [Fe/H].  However we  may use as test values $M_s=10~M_\odot$
for the progenitors of both  PSR~1257+12 and PSR~B1620-26.  As for the
metallicity we assume  0 and -1.05 (the first  is an indicative value,
the  latter is  the average  for  M4 stars),  respectively. Using  the
formule (1), (2) and (3), we find that these planets are very far from
all the other  EPs in the clustering space, and  hence for each pulsar
we have  a single  cluster.  This is  consistent with the  very likely
different origin of the pulsar planets with respect to other EPs.\\
%
%

\acknowledgments

The  author wish to  thank C.~Barbieri  for a  careful reading  of the
paper and financial  support. The paper has been  funded on MIUR grant
PRIN 2006.   Many thanks to  S.~Ortolani and P.~Paolicchi  for helpful
comments and discussion  which improved the quality of  the paper, and
to the  anonymous referee for useful  comments.  I wish  to also thank
N.~Schneider  (on   sabbatical  at  the  University   of  Padova)  for
discussions.  Thanks  to S.~Magrin for IDL support  and M.~Clemens for
corrections to the English.

{}

\clearpage

\clearpage

\clearpage

\clearpage

\clearpage

\begin{deluxetable}{l|lclllccl}
\tabletypesize{\scriptsize}
\tablewidth{0pt}  \tablecaption{Relevant  data  for extrasolar  planet
clusters.   Corr:  All  significant  intracluster  correlations,  i.e.
having 2-tailed probability less then 5\%.  Those of {$\cal C$}{\it2},
{$\cal  C$}{\it  3} and  {$\cal  C$}{\it  4}  are the  most  important
ones. We  show them  in the figs.   \ref{c2}, \ref{c3},  and \ref{c4}.
HT: Hot  Jupiters.  T: Transiting  planets.  MSS: Planets  in multiple
star  systems.   MPS: Multiple  planetary  systems.  SS: Solar  system
planets.}
\tablehead{ \colhead{Cluster} &
\colhead{Prototype}    &   \colhead{Members}   &    \colhead{Corr}   &
\colhead{HJ} & \colhead{T}  & \colhead{MSS} &\colhead{MPS}  & \colhead{SS}} 
\startdata
{$\cal C$}{\it 1}  & HD~41004~A~b   & 11 & $M_p-e$, $M_p-M_s$                       & 2  & -  & 6  & -  &        \\ 
{$\cal C$}{\it 2}  & HD~69830~c     & 46 & $a-e$, $a-$[Fe/H]                        & 17 & 5  & 4  & 13 &        \\ 
{$\cal C$}{\it 3}  & HD~11964~b     & 48 & $a-$[Fe/H], $e-$[Fe/H], $M_s-$[Fe/H]     & 23 & 4  & 11 & 8  & Jupiter  \\ 
{$\cal C$}{\it 4}  & HD~142022~A~b  & 48 & $M_p-e$,  $M_p-M_s$                      & -  & -  & 12 & 12 &    \\ 
{$\cal C$}{\it 5}  & HD~117207~b    & 31 & -                                        & 1  & -  & 7  & 13 &      \\ 
%
\enddata
\label{tab}
\end{deluxetable}   

\clearpage

\begin{deluxetable}{lrrrrrrrrr}
\tabletypesize{\scriptsize}
\tablewidth{0pt}
\tablecaption{Supplementary  Material:  Details  on extrasolar  planet
  clusters.  For each planet  we report  the mass  ($M_J$), semi-major
  axis  (AU), period (days),  eccentricity, stellar  mass ($M_\odot$),
  metallicity, and  the first three principal  components.  Planets in
  bold  are prototypes.  Planets  in multiple  star systems,  multiple
  planet systems, transiting, hot Jupiters are also indicated.}
\tablehead{
\colhead{Planet} & \colhead{$M_p$} & \colhead{$a$} &  \colhead{$P$} & \colhead{$e$} &
\colhead{$M_s$} & \colhead{[Fe/H]} & \colhead{$pc_1$} &
\colhead{$pc_2$} & \colhead{$pc_3$} }
\startdata
\cutinhead{$ Cluster \ \cal  C${\it 1}} 
           HD147513b$^\dag$ &   1.000 &   1.260 &     538.33 &  0.520 &   0.920 &  -0.030 &  -0.426 &  -0.801 &   0.241        \\   
           HD162020b$^{h}$&  13.750 &   0.072 &       8.36 &  0.277 &   0.700 &   0.010 &  -1.171 &  -2.260 &   1.806          \\
          HD178911Bb$^\dag$ &   6.292 &   0.320 &      70.64 &  0.124 &   0.870 &   0.280 &   0.459 &  -0.017 &   1.041        \\   
           HD41004Bb$^{vh,\dag}$&  18.400 &   0.018 &       1.33 &  0.081 &   0.400 &  -0.010 &  -0.883 &  -3.580 &   2.373    \\ 
             HD3651b$^\dag$ &   0.200 &   0.284 &      62.19 &  0.630 &   0.790 &   0.050 &   0.135 &  -0.829 &   1.080        \\   
           HD216770b &   0.650 &   0.460 &     120.08 &  0.370 &   0.900 &   0.230 &   0.646 &   0.260 &   0.774               \\
             HD8574b &   2.230 &   0.760 &     237.07 &  0.400 &   1.040 &  -0.090 &  -0.221 &  -0.714 &  -0.034               \\
           HD114762b$^\dag$ &  11.020 &   0.300 &      65.86 &  0.340 &   0.820 &  -0.500 &  -1.201 &  -3.693 &   0.337        \\   
           HD111232b &   6.800 &   1.970 &    1138.84 &  0.200 &   0.780 &  -0.360 &  -0.678 &  -2.921 &  -0.483               \\
 \bf HD41004Ab$^\dag$ & \bf  2.300 & \bf  1.310 & \bf 653.56 & \bf 0.390 & \bf  0.700 & \bf -0.090 & \bf  0.120 & \bf -1.780 &  \bf 0.368 \\   
            HD65216b &  1.210  &  1.370  &    610.27  & 0.410  &  0.920  & -0.120  & -0.159  & -1.138  & -0.160                \\
\cutinhead{$ Cluster   \ \cal    C${\it 2  }  } 	     	       	         	   	     			       
           HD121504b &   0.890 &   0.320 &      66.09 &  0.130 &   1.000 &   0.160 &   1.306 &   0.358 &   0.092               \\
           HIP14810c$^\ddag$ &   0.951 &   0.458 &     113.73 &  0.281 &   0.990 &   0.230 &   0.727 &   0.542 &   0.484           \\
           HD102117b &   0.140 &   0.149 &      21.55 &  0.060 &   0.950 &   0.180 &   1.885 &   0.373 &   0.084           \\
            HD76700b$^{h}$ &   0.197 &   0.049 &       3.96 &  0.130 &   1.000 &   0.140 &   1.594 &   0.345 &   0.078     \\
           HD209458b$^{h,t}$ &   0.690 &   0.045 &       3.47 &  0.070 &   1.010 &   0.040 &   1.675 &  -0.037 &  -0.208   \\  
             HD2638b$^{h}$ &   0.480 &   0.044 &       3.49 & -0.000 &   0.930 &   0.160 &   2.112 &   0.228 &   0.035     \\
           HD195019b$^\dag$ &   3.670 &   0.138 &      18.16 &  0.014 &   1.060 &   0.080 &   1.114 &   0.043 &  -0.077    \\       
            HD23079b &   2.610 &   1.650 &     737.31 &  0.100 &   1.100 &  -0.240 &   0.124 &  -1.075 &  -1.245           \\
           HD102195b$^{h}$ &   0.488 &   0.049 &       4.12 &  0.060 &   0.926 &   0.096 &   1.906 &  -0.056 &   0.017     \\
           HD330075b$^{h}$ &   0.760 &   0.043 &       3.34 & -0.000 &   0.950 &   0.030 &   2.010 &  -0.243 &  -0.243     \\
             HD4208b &   0.800 &   1.670 &     817.08 &  0.050 &   0.930 &  -0.240 &   0.966 &  -1.432 &  -1.206           \\
            HD27894b &   0.620 &   0.122 &      17.97 &  0.049 &   0.750 &   0.300 &   2.211 &   0.192 &   0.713           \\
             HD4308b &   0.047 &   0.114 &      15.43 & -0.000 &   0.830 &  -0.310 &   2.314 &  -1.889 &  -0.822           \\
            HD37124d$^\ddag$ &   0.600 &   1.640 &     803.93 &  0.140 &   0.910 &  -0.320 &   0.747 &  -1.820 &  -1.183           \\
 \bf HD69830c$^\ddag$ & \bf  0.038 & \bf  0.186 & \bf  31.59 & \bf 0.130 & \bf  0.860 & \bf -0.050 & \bf  1.795 & \bf -0.822 & \bf -0.129   \\
            HD37124b$^\ddag$ &   0.610 &   0.530 &     147.69 &  0.055 &   0.910 &  -0.320 &   1.638 &  -1.760 &  -0.970           \\
            HD69830b$^{h,\ddag}$ &   0.033 &   0.079 &       8.66 &  0.100 &   0.860 &  -0.050 &   1.957 &  -0.809 &  -0.148     \\
           HD189733b$^{vh,t}$ &   1.150 &   0.031 &       2.26 & -0.000 &   0.800 &  -0.030 &   2.205 &  -0.973 &  -0.066  \\  
        OGLE-TR-113b$^{vh,t}$ &   1.320 &   0.023 &       1.43 & -0.000 &   0.780 &   0.140 &   2.223 &  -0.381 &   0.338  \\  
           HD192263b &   0.720 &   0.150 &      23.86 & -0.000 &   0.790 &  -0.200 &   2.236 &  -1.638 &  -0.476           \\
              TrES-1$^{h,t}$ &   0.610 &   0.039 &       3.05 &  0.135 &   0.870 &   0.001 &   1.725 &  -0.636 &   0.067   \\  
           HD117618b &   0.190 &   0.280 &      52.81 &  0.390 &   1.050 &   0.040 &   0.484 &   0.013 &   0.174           \\
           HD130322b &   1.080 &   0.088 &      10.72 &  0.048 &   0.790 &  -0.020 &   2.044 &  -0.976 &   0.034           \\
           HD101930b &   0.300 &   0.302 &      70.46 &  0.110 &   0.740 &   0.170 &   1.980 &  -0.348 &   0.484           \\
           HD168746b$^{h}$ &   0.230 &   0.065 &       6.31 &  0.081 &   0.920 &  -0.070 &   1.878 &  -0.712 &  -0.306     \\
        OGLE-TR-111b$^{h,t}$ &   0.530 &   0.047 &       4.11 & -0.000 &   0.820 &   0.120 &   2.298 &  -0.272 &   0.148   \\  
            HD69830d$^\ddag$ &   0.058 &   0.630 &     196.95 &  0.070 &   0.860 &  -0.050 &   1.755 &  -0.817 &  -0.381           \\
             rhoCrBb &   1.040 &   0.220 &      38.65 &  0.040 &   0.950 &  -0.190 &   1.705 &  -1.148 &  -0.658           \\
            HD37124c$^\ddag$ &   0.683 &   3.190 &    2180.82 &  0.200 &   0.910 &  -0.320 &  -0.316 &  -1.894 &  -1.577           \\
            HD83443b$^{h}$ &   0.410 &   0.040 &       3.29 &  0.080 &   0.790 &   0.330 &   2.122 &   0.443 &   0.769     \\
            HD63454b$^{h}$ &   0.380 &   0.036 &       2.79 & -0.000 &   0.800 &   0.110 &   2.371 &  -0.360 &   0.153     \\
           HD114729b$^\dag$ &   0.820 &   2.080 &    1135.74 &  0.310 &   0.930 &  -0.220 &  -0.148 &  -1.455 &  -0.831    \\       
            HD93083b &   0.370 &   0.477 &     143.79 &  0.140 &   0.700 &   0.150 &   1.840 &  -0.570 &   0.515           \\
             HD6434b &   0.480 &   0.150 &      21.21 &  0.300 &   1.000 &  -0.520 &   0.856 &  -2.327 &  -0.987           \\
              Gl581b$^{h}$ &   0.056 &   0.041 &       5.45 & -0.000 &   0.310 &  -0.330 &   3.299 &  -3.568 &   0.071     \\
          Gliese876d$^{vh,\ddag}$ &   0.023 &   0.021 &       1.94 & -0.000 &   0.320 &  -0.120 &   3.312 &  -2.710 &   0.490    \\
               Gl86b$^\dag$ &   4.010 &   0.110 &      14.96 &  0.046 &   0.790 &  -0.240 &   1.421 &  -2.081 &  -0.168    \\       
            HD99492b$^\dag$ &   0.109 &   0.123 &      17.88 &  0.254 &   0.780 &   0.360 &   1.564 &   0.494 &   1.106    \\       
              GJ436b$^{h}$ &   0.067 &   0.028 &       2.64 &  0.207 &   0.410 &  -0.320 &   2.414 &  -3.291 &   0.293     \\
          Gliese876b$^\ddag$ &   1.935 &   0.208 &      61.00 &  0.025 &   0.320 &  -0.120 &   2.732 &  -2.882 &   0.645           \\
           HIP14810b$^{h,\ddag}$ &   3.840 &   0.069 &       6.67 &  0.148 &   0.990 &   0.230 &   0.796 &   0.359 &   0.633     \\
            HD47536b &   4.960 &   1.610 &     767.70 &  0.200 &   0.940 &  -0.680 &  -0.418 &  -3.520 &  -1.469           \\
          Gliese876c$^\ddag$ &   0.560 &   0.130 &      30.24 &  0.270 &   0.320 &  -0.120 &   2.220 &  -2.848 &   0.987           \\
           HD150706b &   1.000 &   0.820 &     279.61 &  0.380 &   0.940 &  -0.130 &   0.248 &  -1.072 &  -0.107           \\
           HD128311b$^\ddag$ &   2.180 &   1.099 &     469.89 &  0.250 &   0.800 &   0.080 &   0.567 &  -0.742 &   0.352           \\
           HD128311c$^\ddag$ &   3.210 &   1.760 &     951.71 &  0.170 &   0.800 &   0.080 &   0.270 &  -0.820 &   0.083           \\
\cutinhead{$ Cluster   \ \cal    C${\it 3  }  }	     	       	         	   	     				   
           HD188015b$^\dag$ &   1.260 &   1.190 &     456.01 &  0.150 &   1.080 &   0.290 &   0.552 &   1.051 &   0.003    \\       
        OGLE-TR-132b$^{vh,t}$ &   1.190 &   0.031 &       1.68 & -0.000 &   1.350 &   0.430 &   1.225 &   2.522 &  -0.074  \\  
           HD134987b &   1.580 &   0.780 &     245.38 &  0.240 &   1.050 &   0.230 &   0.452 &   0.679 &   0.257           \\
           HD224693b &   0.710 &   0.233 &      35.61 &  0.050 &   1.330 &   0.343 &   1.074 &   2.136 &  -0.239           \\
             Jupiter &   1.000 &   5.203 &    4332.80 &  0.048 &   1.000 &   0.000 &  -1.098 &  -0.397 &  -1.980           \\
            HD73526c$^\ddag$ &   2.500 &   1.050 &     388.67 &  0.140 &   1.020 &   0.280 &   0.516 &   0.732 &   0.227           \\
        GammaCepheib$^\dag$ &   1.430 &   1.940 &     908.07 &  0.200 &   1.180 &   0.000 &  -0.262 &   0.168 &  -0.911     \\      
           HD185269b$^{h}$ &   0.940 &   0.077 &       6.90 &  0.300 &   1.280 &   0.110 &   0.333 &   0.970 &  -0.111      \\
            HD70642b &   2.000 &   3.300 &    2187.59 &  0.100 &   1.000 &   0.160 &  -0.437 &   0.187 &  -0.842            \\
           HD149026b$^{h,t}$ &   0.360 &   0.042 &       2.76 & -0.000 &   1.300 &   0.360 &   1.476 &   2.162 &  -0.209    \\ 
           HD208487b &   0.450 &   0.490 &     109.86 &  0.320 &   1.300 &  -0.060 &   0.093 &   0.387 &  -0.641            \\
            HD27442b$^\dag$ &   1.280 &   1.180 &     427.19 &  0.070 &   1.200 &   0.200 &   0.602 &   1.093 &  -0.531     \\      
            HD88133b$^{h}$ &   0.220 &   0.047 &       3.40 &  0.110 &   1.200 &   0.340 &   1.308 &   1.750 &   0.107      \\
          BD-103166b$^{h}$ &   0.480 &   0.046 &       3.43 &  0.070 &   1.100 &   0.500 &   1.584 &   2.061 &   0.564      \\
            HD68988b$^{h}$ &   1.900 &   0.071 &       6.30 &  0.140 &   1.200 &   0.240 &   0.840 &   1.208 &   0.098      \\
            HD52265b &   1.130 &   0.490 &     117.80 &  0.290 &   1.130 &   0.110 &   0.377 &   0.484 &   0.014            \\
           HD107148b &   0.210 &   0.269 &      48.15 &  0.050 &   1.120 &   0.314 &   1.538 &   1.416 &   0.012            \\
           HD109749b$^{h,\dag}$ &   0.280 &   0.064 &       5.33 &  0.010 &   1.200 &   0.250 &   1.622 &   1.425 &  -0.258 \\    
           HD217107b$^{h,\ddag}$ &   1.330 &   0.073 &       7.13 &  0.132 &   1.020 &   0.370 &   1.320 &   1.213 &   0.614      \\
           HD149143b$^{h}$ &   1.330 &   0.053 &       4.05 &  0.016 &   1.210 &   0.260 &   1.374 &   1.407 &  -0.146     \\
            HD86081b$^{h}$ &   1.500 &   0.039 &       2.56 &  0.008 &   1.210 &   0.257 &   1.373 &   1.384 &  -0.147     \\
           HD160691e$^\ddag$ &   0.522 &   0.921 &     310.59 &  0.067 &   1.080 &   0.280 &   1.134 &   1.109 &  -0.145           \\
             UpsAndb$^{h,\dag,\ddag}$ &   0.690 &   0.059 &       4.59 &  0.029 &   1.300 &   0.130 &   1.285 &   1.222 &  -0.608\\     
           HD104985b &   6.300 &   0.780 &     205.04 &  0.030 &   1.500 &  -0.350 &  -0.661 &  -0.530 &  -1.680           \\
           HD179949b$^{h}$ &   0.950 &   0.045 &       3.08 &  0.022 &   1.280 &   0.220 &   1.305 &   1.495 &  -0.372     \\
              55Cnce$^{h,\dag,\ddag}$ &   0.045 &   0.038 &       2.67 &  0.174 &   1.030 &   0.290 &   1.437 &   1.024 &   0.404\\     
            HD75289b$^{h}$ &   0.420 &   0.046 &       3.52 &  0.054 &   1.050 &   0.290 &   1.729 &   1.094 &   0.184     \\
           HD164922b &   0.360 &   2.110 &    1154.49 &  0.050 &   0.940 &   0.170 &   0.827 &   0.229 &  -0.562           \\
            HD73256b$^{h}$ &   1.870 &   0.037 &       2.53 &  0.030 &   1.050 &   0.290 &   1.517 &   0.983 &   0.274     \\
            HD99109b &   0.502 &   1.105 &     439.84 &  0.090 &   0.930 &   0.315 &   1.235 &   0.772 &   0.169           \\
           HD114783b &   0.990 &   1.200 &     500.34 &  0.100 &   0.920 &   0.330 &   1.068 &   0.754 &   0.248           \\
         OGLE-TR-10b$^{h,t}$ &   0.630 &   0.042 &       2.85 & -0.000 &   1.180 &   0.120 &   1.624 &   0.829 &  -0.470   \\  
           HD160691d$^{h,\ddag}$ &   0.044 &   0.090 &       9.49 & -0.000 &   1.080 &   0.280 &   1.911 &   1.195 &  -0.034     \\
              55Cncb$^{\dag,\ddag}$ &   0.784 &   0.115 &      14.03 &  0.020 &   1.030 &   0.290 &   1.770 &   1.012 &   0.168    \\       
           HAT--P-1b$^{h,t}$ &   0.530 &   0.055 &       4.46 &  0.090 &   1.120 &   0.130 &   1.440 &   0.661 &  -0.197   \\  
              47Umab &   2.600 &   2.110 &    1101.77 &  0.049 &   1.030 &   0.060 &   0.198 &  -0.110 &  -0.747           \\
            HD20367b &   1.070 &   1.250 &     500.32 &  0.230 &   1.040 &   0.100 &   0.346 &   0.170 &  -0.212           \\
            HD10647b &   0.910 &   2.100 &    1074.16 &  0.180 &   1.070 &  -0.030 &   0.025 &  -0.243 &  -0.916           \\
 \bf  HD11964b$^\dag$ & \bf  0.110 & \bf  0.229 & \bf 37.74 & \bf 0.150 &\bf 1.125 &\bf  0.170 &\bf  1.220 &\bf  0.843 &\bf  -0.110    \\       
           HD108874b &   1.360 &   1.051 &     393.31 &  0.070 &   1.000 &   0.140 &   1.016 &   0.239 &  -0.255           \\
            HD16141b$^\dag$ &   0.230 &   0.350 &      75.62 &  0.210 &   1.000 &   0.220 &   1.156 &   0.620 &   0.290    \\       
              51Pegb$^{h}$&   0.468 &   0.052 &       4.21 & -0.000 &   1.060 &   0.160 &   1.873 &   0.629 &  -0.196      \\
           HD212301b$^{vh}$ &   0.450 &   0.036 &       2.43 & -0.000 &   1.050 &   0.180 &   1.905 &   0.679 &  -0.134    \\
           HD187123b$^{h}$&   0.520 &   0.042 &       3.05 &  0.030 &   1.060 &   0.160 &   1.765 &   0.615 &  -0.134      \\
            HD46375b$^{h,\dag}$ &   0.249 &   0.041 &       3.03 &  0.040 &   1.000 &   0.250 &   1.903 &   0.802 &   0.150\\     
            HD49674b$^{h}$ &   0.110 &   0.057 &       4.94 &  0.160 &   1.000 &   0.250 &   1.513 &   0.773 &   0.348     \\
           HD190360c$^{\dag,\ddag}$ &   0.057 &   0.128 &      17.07 &  0.010 &   0.960 &   0.240 &   2.070 &   0.663 &   0.100    \\       
           HD196885b &   1.840 &   1.120 &     383.91 &  0.300 &   1.270 &  -0.390 &  -0.434 &  -1.127 &  -1.387           \\
\cutinhead{$ Cluster   \ \cal    C${\it 4  }  } 	     	       	         	   	     			   
            HD45350b &   1.790 &   1.920 &     961.39 &  0.778 &   1.020 &   0.290 &  -1.989 &   0.590 &   1.043           \\
            HD74156c$^\ddag$ &   6.170 &   3.400 &    2228.53 &  0.583 &   1.050 &   0.130 &  -3.093 &  -0.286 &   0.217           \\
           HD142415b &   1.620 &   1.050 &     386.94 &  0.500 &   1.030 &   0.210 &  -0.556 &   0.440 &   0.632           \\
            HD50554b &   4.900 &   2.380 &    1276.02 &  0.420 &   1.100 &   0.020 &  -1.820 &  -0.374 &  -0.168           \\
  \bf HD142022Ab$^\dag$ & \bf  4.400 &\bf   2.800 & \bf  1716.36 &\bf  0.570 &\bf  0.990 &\bf 0.190 &\bf  -2.245 &\bf  -0.067 & \bf  0.461    \\       
            HD30177b &   9.170 &   3.860 &    2829.03 &  0.300 &   0.950 &   0.250 &  -2.803 &  -0.291 &   0.250           \\
           HD136118b &  11.900 &   2.300 &    1138.96 &  0.370 &   1.240 &  -0.065 &  -3.307 &  -0.836 &  -0.024           \\
            HD39091b &  10.350 &   3.290 &    2069.02 &  0.620 &   1.100 &   0.090 &  -4.114 &  -0.644 &   0.525           \\
             HD4203b &   1.650 &   1.090 &     403.43 &  0.460 &   1.060 &   0.220 &  -0.501 &   0.582 &   0.518           \\
            HD89744b$^\dag$ &   7.990 &   0.890 &     258.49 &  0.670 &   1.400 &   0.180 &  -3.038 &   0.883 &   0.853    \\       
              14Herb &   4.640 &   2.770 &    1770.69 &  0.369 &   0.900 &   0.430 &  -1.411 &   0.645 &   0.784           \\
              55Cncc$^{\dag,\ddag}$ &   0.217 &   0.240 &      42.31 &  0.440 &   1.030 &   0.290 &   0.383 &   0.914 &   0.830    \\       
           HD190228b &   4.990 &   2.310 &    1122.69 &  0.430 &   1.300 &  -0.240 &  -2.217 &  -0.786 &  -1.005           \\
            HD38529c$^{\dag,\ddag}$ &  12.700 &   3.680 &    2177.65 &  0.360 &   1.390 &   0.313 &  -4.434 &   1.004 &   0.094    \\       
           HD160691c$^\ddag$ &   3.100 &   4.170 &    2988.87 &  0.570 &   1.080 &   0.280 &  -2.879 &   0.630 &  -0.078           \\
           HD122430b &   3.710 &   1.020 &     318.75 &  0.680 &   1.390 &  -0.050 &  -2.259 &   0.297 &  -0.013           \\
            HD92788b &   3.860 &   0.970 &     338.35 &  0.270 &   1.060 &   0.240 &  -0.240 &   0.544 &   0.456           \\
           HD213240b$^\dag$ &   4.500 &   2.030 &     954.79 &  0.450 &   1.220 &   0.230 &  -1.855 &   0.853 &   0.188    \\       
           HD217107c$^\ddag$ &   2.500 &   4.410 &    3345.49 &  0.537 &   1.020 &   0.370 &  -2.658 &   0.851 &   0.021           \\
              55Cncd$^{\dag,\ddag}$ &   3.920 &   5.257 &    4330.22 &  0.327 &   1.030 &   0.290 &  -2.716 &   0.496 &  -0.690    \\       
            HD66428b &   2.820 &   3.180 &    1932.63 &  0.465 &   1.146 &   0.310 &  -2.044 &   1.039 &  -0.019           \\
            HD10697b &   6.120 &   2.130 &    1079.77 &  0.110 &   1.100 &   0.150 &  -0.867 &   0.147 &  -0.265           \\
           HD190360b$^{\dag,\ddag}$ &   1.502 &   3.920 &    2891.20 &  0.360 &   0.960 &   0.240 &  -1.481 &   0.313 &  -0.389     \\      
           HD183263b &   3.690 &   1.520 &     631.87 &  0.380 &   1.170 &   0.300 &  -1.076 &   1.079 &   0.389            \\
            HD23596b &   7.190 &   2.720 &    1450.07 &  0.314 &   1.270 &   0.320 &  -2.403 &   1.162 &   0.057            \\
            HD12661b$^\ddag$ &   2.300 &   0.830 &     266.74 &  0.350 &   1.070 &   0.293 &  -0.131 &   0.890 &   0.598            \\
             16CygBb$^\dag$ &   1.680 &   1.680 &     798.74 &  0.689 &   0.990 &   0.080 &  -1.492 &  -0.280 &   0.572     \\      
             HD2039b &   4.850 &   2.190 &    1193.00 &  0.680 &   0.980 &   0.100 &  -2.371 &  -0.507 &   0.731            \\
            HD80606b$^\dag$ &   3.410 &   0.439 &     111.79 &  0.927 &   0.900 &   0.430 &  -1.800 &   0.630 &   2.441     \\      
           HD202206b$^\ddag$ &  17.400 &   0.830 &     255.72 &  0.435 &   1.150 &   0.370 &  -3.670 &   0.160 &   2.124            \\
            HD33564b &   9.100 &   1.100 &     375.61 &  0.340 &   1.250 &  -0.120 &  -1.999 &  -0.743 &  -0.066            \\
            HD20782b$^\dag$ &   1.800 &   1.360 &     578.82 &  0.920 &   1.000 &  -0.050 &  -2.159 &  -0.836 &   0.817     \\      
           HD222582b$^\dag$ &   5.110 &   1.350 &     571.55 &  0.760 &   1.000 &  -0.010 &  -2.285 &  -0.900 &   0.912     \\      
            HD28185b &   5.700 &   1.030 &     382.70 &  0.070 &   0.990 &   0.240 &   0.160 &   0.242 &   0.365            \\
           HD106252b &   6.810 &   2.610 &    1498.43 &  0.540 &   1.050 &  -0.160 &  -2.669 &  -1.439 &  -0.141            \\
           HD168443c$^\ddag$ &  16.870 &   2.840 &    1769.45 &  0.222 &   0.960 &   0.100 &  -3.595 &  -1.429 &   0.809           \\
           HIP75458b &   8.820 &   1.275 &     511.15 &  0.712 &   1.050 &   0.030 &  -2.934 &  -0.878 &   1.179           \\
           HD141937b &   9.700 &   1.520 &     681.35 &  0.410 &   1.000 &   0.160 &  -2.115 &  -0.501 &   0.991           \\
            HD33636b &   9.280 &   3.560 &    2454.89 &  0.530 &   0.990 &  -0.130 &  -3.547 &  -1.735 &  -0.083           \\
            HD74156b$^\ddag$ &   1.860 &   0.294 &      56.78 &  0.636 &   1.050 &   0.130 &  -0.702 &   0.145 &   0.947           \\
     EpsilonEridanib &   1.550 &   3.390 &    2500.25 &  0.702 &   0.830 &  -0.100 &  -2.158 &  -1.523 &  -0.070           \\
           HD210277b &   1.230 &   1.100 &     439.06 &  0.472 &   0.920 &   0.190 &  -0.208 &   0.063 &   0.682           \\
              70Virb &   7.440 &   0.480 &     115.45 &  0.400 &   1.100 &  -0.030 &  -1.246 &  -0.716 &   0.545           \\
            HD37605b &   2.300 &   0.250 &      50.98 &  0.677 &   0.800 &   0.390 &  -0.435 &   0.346 &   2.048           \\
           HD168443b$^\ddag$ &   7.480 &   0.290 &      58.00 &  0.530 &   0.960 &   0.100 &  -1.332 &  -0.679 &   1.357           \\
             GJ3021b$^\dag$ &   3.320 &   0.490 &     131.83 &  0.505 &   0.900 &   0.200 &  -0.383 &  -0.125 &   1.185    \\       
           HD154857b &   1.800 &   1.110 &     394.62 &  0.510 &   1.170 &  -0.230 &  -0.945 &  -0.874 &  -0.506           \\
            HD81040b &   6.860 &   1.940 &    1003.92 &  0.526 &   0.960 &  -0.160 &  -2.103 &  -1.695 &   0.216           \\
 \cutinhead{$ Cluster   \ \cal    C${\it   5  }}	     	       	         	   	     
           HD160691b$^\ddag$ &   1.670 &   1.500 &     645.23 &  0.310 &   1.080 &   0.280 &  -0.248 &   0.915 &   0.205           \\
           HD202206c$^\ddag$ &   2.440 &   2.550 &    1385.58 &  0.267 &   1.150 &   0.370 &  -0.951 &   1.403 &  -0.085           \\
            HD11977b &   6.540 &   1.930 &     707.49 &  0.400 &   1.910 &  -0.210 &  -3.337 &   1.104 &  -1.800           \\
             UpsAndd$^{\dag,\ddag}$ &   3.950 &   2.510 &    1272.09 &  0.242 &   1.300 &   0.130 &  -1.444 &   0.808 &  -0.739    \\       
           HD196050b$^\dag$ &   3.000 &   2.500 &    1374.86 &  0.280 &   1.100 &   0.000 &  -1.017 &  -0.251 &  -0.671    \\       
            HD40979b$^\dag$ &   3.320 &   0.811 &     256.33 &  0.230 &   1.080 &   0.194 &   0.054 &   0.488 &   0.258    \\       
           HD187085b &   0.750 &   2.050 &     970.36 &  0.470 &   1.220 &   0.050 &  -1.174 &   0.450 &  -0.491           \\
            HD33283b &   0.330 &   0.168 &      22.58 &  0.480 &   1.240 &   0.366 &  -0.116 &   1.839 &   0.725           \\
           HD108147b &   0.400 &   0.104 &      10.87 &  0.498 &   1.270 &   0.200 &  -0.223 &   1.270 &   0.390           \\
           HD169830c$^\ddag$ &   4.040 &   3.600 &    2105.72 &  0.330 &   1.400 &   0.210 &  -2.532 &   1.360 &  -0.941           \\
           HD177830b &   1.280 &   1.000 &     337.51 &  0.430 &   1.170 &   0.000 &  -0.489 &   0.101 &  -0.185           \\
            HD82943c$^\ddag$ &   2.010 &   0.746 &     219.28 &  0.359 &   1.150 &   0.270 &  -0.204 &   1.069 &   0.428           \\
           HD216437b &   2.100 &   2.700 &    1565.15 &  0.340 &   1.070 &   0.000 &  -1.091 &  -0.295 &  -0.657           \\
            HD12661c$^\ddag$ &   1.570 &   2.560 &    1445.35 &  0.200 &   1.070 &   0.293 &  -0.407 &   0.948 &  -0.306           \\
            HD89307b &   2.730 &   4.150 &    2737.38 &  0.270 &   1.270 &  -0.230 &  -2.148 &  -0.653 &  -2.027           \\
            HD73526b$^\ddag$ &   2.900 &   0.660 &     193.66 &  0.190 &   1.020 &   0.280 &   0.474 &   0.694 &   0.481           \\
  \bf HD117207b & \bf  2.060 &\bf   3.780 &\bf  2629.78 & \bf 0.160 &\bf   1.040 &\bf   0.270 &\bf  -0.981 &\bf   0.702 &\bf  -0.730   \\
              HD142b$^\dag$ &   1.000 &   0.980 &     337.72 &  0.380 &   1.100 &   0.040 &  -0.119 &   0.083 &  -0.089    \\       
            HD50499b &   1.710 &   3.860 &    2456.46 &  0.230 &   1.270 &   0.230 &  -1.613 &   1.257 &  -1.146           \\
           HD169830b$^\ddag$ &   2.880 &   0.810 &     224.83 &  0.310 &   1.400 &   0.210 &  -0.711 &   1.546 &  -0.164           \\
           HD216435b &   1.490 &   2.700 &    1448.61 &  0.340 &   1.250 &   0.150 &  -1.283 &   0.898 &  -0.717           \\
            HD19994b$^\dag$ &   2.000 &   1.300 &     465.64 &  0.200 &   1.350 &   0.230 &  -0.327 &   1.566 &  -0.472    \\       
            HD82943b$^\ddag$ &   1.750 &   1.190 &     441.84 &  0.219 &   1.150 &   0.270 &   0.087 &   1.124 &   0.008           \\
            HD38529b$^{\dag,\ddag}$ &   0.780 &   0.129 &      14.35 &  0.290 &   1.390 &   0.313 &   0.185 &   2.120 &   0.066    \\       
              HR810b &   1.940 &   0.910 &     300.71 &  0.240 &   1.110 &   0.250 &   0.199 &   0.908 &   0.183           \\
           HD108874c$^\ddag$ &   1.018 &   2.680 &    1601.77 &  0.250 &   1.000 &   0.140 &  -0.412 &   0.158 &  -0.498           \\
             UpsAndc$^{\dag,\ddag}$ &   1.980 &   0.830 &     242.07 &  0.254 &   1.300 &   0.130 &  -0.168 &   1.017 &  -0.342    \\       
           HD118203b$^{h}$ &   2.130 &   0.070 &       6.09 &  0.309 &   1.230 &   0.100 &   0.151 &   0.676 &   0.081     \\
              47Umac$^\ddag$ &   1.340 &   7.730 &    7730.20 & -0.000 &   1.030 &   0.060 &  -2.425 &  -0.157 &  -2.795           \\
            HD62509b &   2.900 &   1.690 &     587.98 &  0.020 &   1.860 &   0.190 &  -1.041 &   2.954 &  -1.817           \\
            HD72659b &   2.960 &   4.160 &    3174.99 &  0.200 &   0.950 &  -0.140 &  -1.372 &  -1.282 &  -1.390           \\
\enddata	
\tablenotetext{\dag}{Planets in multiple star systems.}
\tablenotetext{\ddag}{Planets in multiple planetary systems.}
\tablenotetext{h}{Hot Jupiters.}
\tablenotetext{vh}{Very hot Jupiters.}
\tablenotetext{t}{Transiting planets.}
\label{tab_sup}
\end{deluxetable}

\end{document}